\documentclass[fleqn,usenatbib]{mnras}

\usepackage{newtxtext}
\usepackage{mathptmx}
\usepackage{comment}

\usepackage[T1]{fontenc}

\DeclareRobustCommand{\VAN}[3]{#2}
\let\VANthebibliography\thebibliography
\def\thebibliography{\DeclareRobustCommand{\VAN}[3]{##3}\VANthebibliography}

\usepackage{graphicx}	
\usepackage{amssymb}	
\usepackage{amsmath}	

\def\araa{ARA\&A}       	
\def\apj{ApJ}           	
\def\apjl{ApJ}          	
\def\apjs{ApJS}         	
\def\aap{A\&A}          	
\def\mnras{MNRAS}       	
\def\pasj{PASJ}         	
\def\nat{Nature}        	

\title[Capturing inside-out quenching]{
Capturing the inside-out quenching by black holes with far-infrared atomic line ratios}

\author[S. Inoue et al.]{
{Shigeki Inoue$^{1,2}$\thanks{E-mail: inouesg@ccs.tsukuba.ac.jp}, Hiroshi Matsuo$^{3}$, Naoki Yoshida$^{4,5,6}$, Hidenobu Yajima$^{1}$ \& Kana Moriwaki$^{4}$}
\\\
$^{1}$Center for Computational Sciences, University of Tsukuba, Ten-nodai, 1-1-1 Tsukuba, Ibaraki 305-8577, Japan\\
$^{2}$Chile Observatory, National Astronomical Observatory of Japan, Mitaka, Tokyo 181-8588, Japan\\
$^{3}$Advanced Technology Center, National Astronomical Observatory of Japan, Mitaka, Tokyo 181-8588, Japan\\
$^{4}$Department of Physics, School of Science, The University of Tokyo, Bunkyo, Tokyo 113-0033, Japan\\
$^{5}$Research Center for the Early Universe, School of Science, The University of Tokyo, Bunkyo, Tokyo 113-0033, Japan\\
$^{6}$Kavli Institute for the Physics and Mathematics of the Universe (WPI), UTIAS, The University of Tokyo, Chiba 277-8583, Japan\\
}

\date{Accepted XXX. Received YYY; in original form ZZZ}

\pubyear{2020}

\begin{document}
\label{firstpage}
\pagerange{\pageref{firstpage}--\pageref{lastpage}}
\maketitle

\begin{abstract}
We propose to use relative strengths of far-infrared fine structure lines from galaxies to characterise early phases of the inside-out quenching by massive black holes (BHs). The BH feedback is thought to quench star formation by evacuating the ambient gas. In order to quantify the feedback effect on the gas density in the galactic centres, we utilise the outputs of IllustrisTNG and Illustris simulations, which implement different BH feedback models. We devise a physical model of H$_{\rm ~II}$ regions and compute the intensities of [O$_{\rm ~III}$] $52$ and $88~{\rm \mu m}$ lines. The line intensity ratio is sensitive to the local electron density, and thus can be used to measure the strength and physical extent of the BH quenching. If the BH feedback abruptly operates and expel the gas when it grows to a certain mass, as modelled in IllustrisTNG, the low-density gas yields relatively weak [O$_{\rm ~III}$] $52$ line with respect to $88~{\rm \mu m}$. In contrast, if the feedback strength and hence the local gas density are not strongly correlated with the BH mass, as in Illustris, the line ratio is not expected to vary significantly among galaxies with different evolutionary stages. We find these features are reproduced in the simulations. We also show that the line ratios are not sensitive to the aperture size for measurement, and thus observations do not need to resolve the galactic centres. We argue that the integrated line ratios can be used to capture the onset of the inside-out quenching by BHs.
\end{abstract}

\begin{keywords}
methods: numerical -- galaxies: evolution -- galaxies: nuclei -- quasars: supermassive black holes
\end{keywords}



\section{Introduction}
\label{Intro}
A wide variety of physical processes occur in the central regions of galaxies through the co-evolution of galaxies and massive black holes (BHs). Detailed observations of galactic inner regions can therefore provide important clues to the evolutionary stages of the galaxies. It is theoretically expected that, in massive galaxies in the high-redshift Universe, a large amount of gas is funnelled into the galactic centre, where the earliest episode of star formation should have taken place. There, the central BH can grow by accreting a large amount of gas, but the details of this process have not been elucidated \citep[e.g.][and references therein]{ivh:20}. When there is a large amount of dense gas in the galactic centre, a massive BH can outshine as a quasar by accreting the ambient gas at a high rate. When the gas around a BH is diffuse and the accretion rate is low, it can expel the gas from the galactic centre through various processes such as high-energy radiation and launching a jet. If the feedback effect of active galactic nuclei (AGN) is strong enough to expel most of the gas from the galaxy, star formation activity there is effectively `quenched', and the galaxy evolves to a quiescent galaxy that consists of old and red stars with little gas. Hence the AGN feedback plays a key role in the evolution of galaxies that host massive BHs.

It is known that the star formation efficiencies of galaxies, i.e. ratios of stellar to halo masses, peaks at a halo mass of $\sim10^{12}~{\rm M_\odot}$, and declines toward more massive galaxies \citep[e.g.][]{sm:12,bwh:19}. This trend is often attributed to the feedback by massive BHs. In a colour-magnitude diagram, quenched galaxies are generally found in a `red sequence' as elliptical and lenticular galaxies, which is distinct from a `blue cloud' consisting of star-forming galaxies. Several quenching mechanisms have been proposed to transform galaxies from blue to red through the so-called `green valley' \citep[e.g.][]{mbt:09,sus:14,jpn:20}, and the BH feedback is thought to be a major process that shapes the colour bimodality of the local galaxies by quenching star formation in massive galaxies.

Because the AGN feedback is driven by the central BH, star-formation quenching is expected to proceed from the galactic centre first and then propagate outwards. This "inside-out" quenching process would begin with lowering the gas densities in the central regions, whereas the outer regions still sustain their star formation activity with a sufficient amount of dense gas. Contrastingly, if there are no massive BHs are or the feedback is insufficient, galaxies usually have gas density profiles increasing towards the centre. Therefore, measuring the gas density at the galactic centre may provide a probe of the inside-out quenching by massive BHs. Namely, if a star-forming galaxy is observed to have a low gas density at the centre, it may be a sign of the onset of the BH quenching of star formation. 

Although it is difficult to measure directly the gas densities by observations especially for distant galaxies, one can use the intensity ratio between a specific pair of emission lines to estimate the local electron density. In the present study, we focus on the fine structure lines of doubly ionized oxygen at the wavelengths of $52$ and $88~{\rm \mu m}$ (hereafter [O$_{\rm ~III}]52$ and [O$_{\rm ~III}]88$). The relative strength of [O$_{\rm ~III}]52$ to $88$ (hereafter [O$_{\rm ~III}]52/88$) is sensitive to the electron density of the line emitting region owing to their different critical densities \citep[e.g.][]{ka:88}. The [O$_{\rm ~III}$] lines are detected in various astronomical objects including low- and high-redshift galaxies \citep[e.g.][]{dcl:14,itm:16,nbd:19}. 
We apply a physical model of the [O$_{\rm ~III}$] line emission to the output of cosmological hydrodynamics simulations and examine whether or not and how the BH feedback affects the line ratio.

Since the detailed physics of BH formation and growth are not know, and numerical simulations do not have a sufficient resolution to reproduce the interaction between BHs and the surrounding gas, even the state-of-the-art cosmological simulations treat the formation, growth and feedback of BHs as sub-resolution models (see Section \ref{BHmodel}). Unfortunately, there is not a well-established model of BH feedback, and recent simulations adopt different numerical implementation, especially for massive BHs. The major purposes of our study are to predict the line ratios of [O$_{\rm ~III}]52/88$ for star-forming galaxies using cosmological simulations and to show how characteristic features of BH feedback models are inferred from the line ratio.

We utilise the outputs of the cosmological simulation of the Next Generation Illustris (hereafter IllustrisTNG or TNG). \citet{nps:18} show that IllustrisTNG reproduces well the colour bimodality of galaxies observed in the local Universe thanks to its improved model for BH feedback in massive galaxies. However, \citet{hsc:20} postprocess the simulation data and model sub-millimetre galaxies (SMGs) at redshift $z=2$, to show that TNG significantly underestimates the SMG luminosity function. They argue that the BH quenching in TNG appears to operate too early. It is thus important to study in detail the feedback effects caused by the massive BHs in the simulations. Using the two sets of simulations allows us to make a quantitative comparison; we apply the same analysis to the original Illustris simulation (hereafter Illustris), in which the BH feedback model is different from that of TNG. \citet{hsc:20} also shows that Illustris can reproduce the luminosity function of SMGs better than TNG. With our postprocessing model of the [O$_{\rm ~III}$] lines, we can address how the differences in the BH feedback models affect the line ratios. It is worth noting here that we do not aim at judging which simulation is more realistic. We propose a method or diagnostics to constrain BH models using observations of emission lines from star-forming galaxies at low and high redshifts. The results of our study would eventually help to understand the behaviours of BH quenching observationally, such as the dependence of the feedback efficiency on BH mass and on other physical parameters in the real Universe.

Because far-infrared emission of atomic fine-structure lines is optically thin, their relative strengths can be used as a direct probe of the gas (electron) density of the emitting region. High-redshift quasars at $z\gtrsim6$ are bright in the line emission such as [O$_{\rm ~III}$] and [N$_{\rm ~II}$] and have been observed with Atacama Large Millimetre/submillimetre Array (ALMA). Our method using these line pairs could be applicable to such quasars in the early Universe. However, because there are no massive BHs at $z\gtrsim6$ in the above simulations, we demonstrate our method with snapshot data at $z\leq4$.

In Section \ref{sims}, we describe the cosmological simulations of IllustrisTNG and Illustris. In Section \ref{modellinglines}, we explain our models to compute the [O$_{\rm ~III}$] lines from snapshot data of the simulations. We present our results in Section \ref{Res} and discuss them in Section \ref{dis}. We summarise this study and draw our conclusions in Section \ref{con}.

\section{The cosmological simulations}
\label{sims}
We utilise publicly available data sets of IllustrsTNG and Illustris. The details of the simulations are presented on the respective web sites\footnote{https://www.tng-project.org/}$^,$\footnote{https://www.illustris-project.org/} and in related papers including \citet{TNG}, \citet{wsh:17} and \citet{psn:18} for IllustrisTNG, and \citet{vgs:14a,vgs:14b}, \citet{gvs:14} and \citet{svg:15} for Illustris. Both simulations are performed with an $N$-body/moving-mesh hydrodynamics code {\sc Arepo} \citep{arepo,wvr:20}. The sub-resolution physics such as gas cooling, star formation and supernovae implemented are essentially the same in the simulations. The main differences between the TNG and Illustris are in their BH feedback models and implementation of magneto-hydrodynamics (see Section \ref{BHmodel}). Although only TNG includes the effects of magnetic fields, \citet{pgg:17} show that the presence of magnetic fields hardly affects the formation history of the simulated galaxies. This study focuses on the runs of TNG100-1 and Illustris-1 for the IllustrisTNG and Illustris simulations, respectively. Their resolutions and sizes of the simulated volumes are hardly different. Their simulation boxes have comoving side lengths of $\simeq110~{\rm Mpc}$, and the mass-resolutions for dark matter and gas are $\simeq7$ and $1.4\times10^6~{\rm M_\odot}$. Hence, the comparison between TNG100-1 and Illustris-1 is expected to extract physical influences by their different BH feedback models. 

In the simulations, dense gas cells with $\rho_{\rm cell}>n_{\rm H,SF}=0.1~{\rm cm^{-3}}$ are converted to stellar particles according to a stochastic model of star formation. The mass-resolution of a stellar particle is therefore comparable to that of the parent gas cell. Star formation rate (SFR) is calculated as
\begin{equation}
\dot{m}_{\rm star} = f_{\rm M} \frac{m_{\rm cell}}{t_{\rm SF}}
\label{cell_sfr}
\end{equation}
where $m_{\rm cell}$ is a mass of the parent cell, $f_{\rm M}$ is the mass fraction of cold gas computed with a model of \citet[][see also \citealt{sh:03}]{ykk:97}, and the star-formation time-scale $t_{\rm SF}$ is approximated as a free-fall time within the cell: $t_{\rm SF}\equiv1/\sqrt{G\rho_{\rm cell}}$. A stellar particle is assumed to have the initial mass function (IMF) of \citet{c:05}. Type-II supernovae (SNe) are triggered immediately following the star formation, and a feedback model of \citet{sh:03} is adopted to represent stellar feedback effects. Type-Ia SNe and asymptotic giant branch stars eject mass and metals into nearby gas cells. 

Gravitationally bound structures are identified with the friend-of-friend and {\sc SUBFIND} grouping algorithms \citep[e.g.][]{swt:01}. In this study, the total masses and SFRs are computed for each {\sc SUBFIND} group (galaxy). When a single galaxy hosts multiple BHs, we define the most massive one to be the representative BH of the galaxy and consider its mass to be the BH mass, $M_{\rm BH}$, of the galaxy. 

\subsection{The black hole models}
\label{BHmodel}
The details of the BH model of IllustrisTNG and Illustris are described in \citet{wsh:17} and \citet{ssd:07,svg:15}, respectively. Briefly, BHs are seeded with the initial mass of $M_{\rm seed}=1.2\times10^6~{\rm M_\odot}$ when their friend-of-friend host haloes reach $M_{\rm FOF}=7.4\times10^{10}~{\rm M_\odot}$ in TNG, whereas the corresponding masses are
$M_{\rm seed}=1.4\times10^5~{\rm M_\odot}$ and $M_{\rm FOF}=7.1\times10^{10}~{\rm M_\odot}$ in Illustris. The seed BHs are placed at the potential centres of their host haloes and move together with them. The BHs increase their masses by accreting the surrounding gas at rates $\dot{M}_{\rm Bondi}$ given by the Bondi–Hoyle–Lyttleton formula limited by the Eddington rate $\dot{M}_{\rm Edd}$, i.e. $\dot{M}_{\rm BH}=\min(\dot{M}_{\rm Bondi},\dot{M}_{\rm Edd})$. In computing $\dot{M}_{\rm Bondi}$, IllustrisTNG does not use an artificial boost factor nor take into account relative velocity between the BH and ambient gas unlike Illustris.  

In both the simulations, the BH model assumes two feedback modes according to whether the Eddington ratios $\dot{M}_{\rm Bondi}/\dot{M}_{\rm Edd}$ are higher or lower than a critical value $\chi$. An important point in our study is that TNG and Illustris adopt different definitions of $\chi$. IllustrisTNG assumes $\chi$ to be a function of $M_{\rm BH}$ as 
\begin{equation}
\chi=\min\left[\chi_0\left(\frac{M_{\rm BH}}{10^8~{\rm M_\odot}}\right)^\beta,0.1\right],
\label{criterion_TNG}
\end{equation}
where $\chi_0=0.002$ and $\beta=2.0$, whereas Illustris sets a constant value of $\chi=0.05$. When $\dot{M}_{\rm Bondi}/\dot{M}_{\rm Edd}>\chi$, an inefficient thermal feedback mode called "quasar mode" is switched on in both simulations. When $\dot{M}_{\rm Bondi}/\dot{M}_{\rm Edd}<\chi$, the BH feedback model largely differ between TNG and Illustris.
In TNG, the criterion given by equation (\ref{criterion_TNG}) quadratically increases with $M_{\rm BH}$ until $M_{\rm BH}$ reaches $10^{8.85}~{\rm M_\odot}$.\footnote{When $M_{\rm BH}>10^{8.85}~{\rm M_\odot}$, $\chi$ is limited to 0.1 according to equation (\ref{criterion_TNG}).} On the other hand, the Eddington ratio scales as $\dot{M}_{\rm Bondi}/\dot{M}_{\rm Edd}\propto M_{\rm BH}$ if the properties of ambient gas do not change. The transition of the BH feedback is thus closely linked to $M_{\rm BH}$ in TNG.\footnote{As long as $M_{\rm BH}<10^{8.85}~{\rm M_\odot}$, a BH growing at a constant Eddington ratio necessarily reaches the condition of $\dot{M}_{\rm Bondi}/\dot{M}_{\rm Edd}<\chi$ at a certain $M_{\rm BH}$, when the BH feedback is switched to the low-accretion mode.} This modelling makes $M_{\rm BH}$ a determinant quantity to switch between the two feedback modes, and massive BHs with $M_{\rm BH}\gtrsim10^{8.5}~{\rm M_\odot}$ are mostly in the low-accretion mode (see Section \ref{SFMS}).

In the quasar mode with $\dot{M}_{\rm Bondi}/\dot{M}_{\rm Edd}>\chi$, a BH distributes thermal energies to nearby gas cells, given as $\dot{E}_{\rm qsr}=\epsilon_{\rm qsr}L_{\rm BH}$, where the thermal coupling efficiencies are $\epsilon_{\rm qsr}=0.1$ and $0.05$ in IllustrisTNG and Illustris, respectively. 
The bolometric luminosity is given by $L_{\rm BH}=\epsilon_{\rm r}\dot{M}_{\rm BH}c^2$ with radiative efficiency $\epsilon_{\rm r}=0.2$, where $c$ is the speed of light. This high-accretion mode occurs essentially in the same manner between the two simulations.

In TNG, a BH in the low-accretion mode with $\dot{M}_{\rm Bondi}/\dot{M}_{\rm Edd}<\chi$ injects momenta to nearby gas cells \citep[`kinetic mode',][]{wsh:17}. The energy-injection rate in this mode is given as $\dot{E}_{\rm kin}=\epsilon_{\rm kin}\dot{M}_{\rm BH}c^2$ with the efficiency $\epsilon_{\rm kin}=0.2$.\footnote{If gas around a BH has a hydrogen number density lower than $n_{\rm H}=10^{-3}~{\rm cm^{-3}}$, the efficiency scales linearly with the local density as $\epsilon_{\rm kin}=\min(20n_{\rm H}/n_{\rm H,SF},0.2)$.} The injection per time-step $\dot{E}_{\rm kin}\Delta t$ is accumulated in time, and the integrated energy is released every time it exceeds a certain threshold given by equation (13) of \citet{wsh:17}. The accumulation makes the BH feedback strong enough to blow out the gas around the BH to halo regions. Since the kinetic energy injected as momentum is not converted quickly to thermal energy that is lost by gas cooling, the kinetic mode can cause efficient feedback in galaxies with massive BHs. 

In Illustris, the low-accretion mode with $\dot{M}_{\rm Bondi}/\dot{M}_{\rm Edd}<\chi$ injects thermal energies to gas within a `bubble' region in the galaxy \citep[`radio mode',][]{ssd:07}. The injected thermal energy into the bubble is given as $E_{\rm rad}=\epsilon_{\rm rad}\delta_{\rm BH}M_{\rm BH}c^2$, where $\epsilon_{\rm rad}=0.35$. The injection is operated every time the BH increments its mass by $\delta_{\rm BH}=1~{\rm per~cent}$ from $M_{\rm BH}$ at the previous injection. The bubble is randomly located within a certain distance from a BH, and therefore the thermal energy is injected effectively by avoiding the high-density region at the galactic centre where the injected energy can be quickly radiated away.

The low-accretion modes are the main driver of the quenching process in massive galaxies in both the simulations although their models are largely different. We expect that characteristic features of the distinct BH feedback models can be found in galaxies that host massive BHs while still forming stars actively. We focus on the early stage of the BH quenching occurring in massive, star-forming galaxies.

\section{[O$_{\rm ~III}]$ line model}
\label{modellinglines}
It is well known that a combination of far-infrared emission lines can be used to estimate the local election density of the line emitting region. Although we consider specifically the line ratio of [O$_{\rm ~III}]52/88$ in this study, other pairs such as [N$_{\rm ~II}$] lines at $122$ and $205~{\rm \mu m}$ \citep[e.g.][]{zlx:16,dgi:20} can also be the density indicator (see also Section \ref{dis_obs}). Far-infrared fine structure lines are useful to observationally study the density structure of high-redshift galaxies, in which the inside-out quenching driven by massive BHs can be occuring.

\subsection{Method}
\label{method}
Because the [O$_{\rm ~III}]$ lines are mainly emitted from H$_{\rm ~II}$ regions around young massive stars, we devise a physical model of H$_{\rm ~II}$ regions and implement in each gas cell in the simulations. We first evaluate the following physical quantities; a typical lifetime of a H$_{\rm ~II}$ region $t_{\rm HII}$ and the characteristic mass of a star cluster $M_{\rm cl}$. We assume that the star cluster is the dominant radiation source that forms the H$_{\rm ~II}$ region. Note that $M_{\rm cl}$ is defined as the initial mass of a cluster. The number of H$_{\rm ~II}$ regions within a single gas cell is estimated as
\begin{equation}
N_{\rm HII}=\frac{\dot{m}_{\rm star}t_{\rm HII}}{M_{\rm cl}},
\label{Nhii}
\end{equation}
where $\dot{m}_{\rm star}$ is an SFR of a cell, given by equation (\ref{cell_sfr}). We set the parameters to be $t_{\rm HII}=1~{\rm Myr}$ \citep[e.g.][]{k:17,fys:20} and $M_{\rm cl}=10^5~{\rm M_\odot}$. We allow $N_{\rm HII}$ to be less than 1, i.e. a computational cell covers only a fraction of a star-forming region. We assume that $\dot{m}_{\rm star}$ is constant during $t_{\rm HII}$, and that the cluster stellar population is represented by time-integrated \citeauthor{c:05} IMF evolved over $t_{\rm HII}$. We consider the metallicity of the stars to be the same as that of its parent gas cell. From these quantities, we compute the spectral energy distribution (SED) of a star cluster using {\sc P\'{E}GASE.2} \citep{PEGASE2}, and calculate the production rate of ionizing photons, $\dot{N}_{\rm ph}$, with energies higher than $13.6~{\rm eV}$.

Since the spatial resolutions of star-forming gas cells with $\rho_{\rm cell}>n_{\rm H,SF}=0.1~{\rm cm^{-3}}$ are in the order of $\sim10$--$100~{\rm pc}$ in the runs of TNG100-1 and Illustris-1, star-forming regions are not resolved in the simulations. The actual physical density in a star-forming region is therefore expected to be significantly higher than the cell density $\rho_{\rm cell}$. We adopt a model for interstellar matter (ISM) used in \citet{iyy:20}, where a gas cell is considered to consist of cold and warm neutral media (CNM and WNM) with their density contrast of $\rho_{\rm CNM}/\rho_{\rm WNM}=100$ \citep{wmh:95}. The ISM model considers that the CNM and WNM are in pressure equilibrium due to the balance between (unresolved) SN feedback and thermal instability by cooling \citep{sh:03}. According to the ISM model, we can compute a density and volume of the CNM for each gas cell. The density enhancement factors are estimated to be $\rho_{\rm CNM}/\rho_{\rm cell}\sim2$--$20$ depending on cell density \citep[see figure 1 of ][]{iyy:20}. Although the CNM and WNM are unresolved in the simulations, stars are actually expected to form in the CNM. We therefore consider that all H$_{\rm ~II}$ regions reside in CNM, and the gas density inside the H$_{\rm ~II}$ regions is approximated to be $\rho_{\rm CNM}$ for each cell.

To compute the line emission of [O$_{\rm ~III}]$ based on the above model for H$_{\rm ~II}$ regions, we use a method described in \citet{mys:18}. Assuming the density within an H$_{\rm ~II}$ region to be uniform, a Str\"{o}mgren sphere forming the H$_{\rm ~II}$ region has a radius of 
\begin{equation}
r_{\rm S}=\left(\frac{3\dot{N}_{\rm ph}}{4\pi n_{\rm e}^2\alpha_{\rm B}}\right)^\frac{1}{3},
\label{StromgrenRadius}
\end{equation}
where $\alpha_{\rm B}$ is the case-B hydrogen recombination coefficient, and we assume a constant value of $\alpha_{\rm B}=2.6\times10^{-13}~{\rm cm^3~s^{-1}}$.\footnote{Although the coefficient $\alpha_{\rm B}$ actually depends on temperature of free electrons, we adopt the value estimated at $10^4~{\rm K}$.} Considering a fully ionized state, the electron number density $n_{\rm e}$ is equal to the hydrogen number density estimated for the H$_{\rm ~II}$ region by the above ISM model: $\rho_{\rm CNM}$. In all snapshots of both simulations, we confirm that the total volume of the Str\"{o}mgren spheres, $4\pi r_{\rm S}^3N_{\rm HII}/3$, does not exceed a volume of CNM in any star-forming cell with $\rho_{\rm cell}>n_{\rm H,SF}$ with the above parameter settings. Eventually, the volume-averaged ionization parameter inside $r_{\rm S}$ is given as 
\begin{equation}
\overline{U}=\frac{3}{4c}\left(\frac{3\dot{N}_{\rm ph}n_{\rm e}\alpha_{\rm B}^2}{4\pi}\right)^\frac{1}{3}.
\label{IonizationParam}
\end{equation}
In the plane-parallel case, the ionization parameter at an inner surface of the H$_{\rm ~II}$ region becomes $U_{\rm in}=2\overline{U}$. 

For the H$_{\rm ~II}$ region modelled above, we adopt a spectral synthesis code {\sc Cloudy} \citep[version 17.02 of the code last described in][]{cloudy} designed to simulate conditions in ISM. For our {\sc Cloudy} model, we input an SED of the radiation source, ionization parameter $U_{\rm in}$, density $n_{\rm e}$, metallicity $Z$ of gas, and redshift $z$ that sets the temperature of the cosmic microwave background. The SED is computed with {\sc P\'{E}GASE.2} for our star cluster model. For the IllustrisTNG simulation, we use an oxygen-based metallicity: $Z\equiv Z_\odot y_{\rm O}/y_{\rm O,\odot}$ where $Z_\odot$ and $y_{\rm O,\odot}$ are the solar metallicity and oxygen abundance. This treatment is motivated by the fact that the [O$_{\rm ~III}]$ lines are more sensitive to oxygen abundance than the total metallicity. Since the original Illustris simulation does not have information on oxygen abundance, we use a metallicity assigned to a gas cell as a proxy. At each redshift $z$, we generate a look-up table of line intensities of [O$_{\rm ~III}]52$, [O$_{\rm ~III}]88$ and H$\alpha$ as a function of the three parameters of $U_{\rm in}$, $n_{\rm e}$ and $Z$ in logarithmic spacing of $0.48~{\rm dex}$, although $U_{\rm in}$ depends only on $n_{\rm e}$ at a given $Z$ (see below). The line emission of H$\alpha$ is used for parameter calibration (see below). The look-up table covers the whole parameter space that the star-forming gas cells in the simulations can have. We estimate the line intensities of each gas cell from the tabulated values multiplied by $N_{\rm HII}$. We assume that low-density cells with $\rho_{\rm cell}<n_{\rm H,SF}$ do not emit any radiation since they do not form stars. In this study, we do not take into account dust attenuation.

\begin{figure}
  \includegraphics[bb=0 0 431 286, width=\hsize]{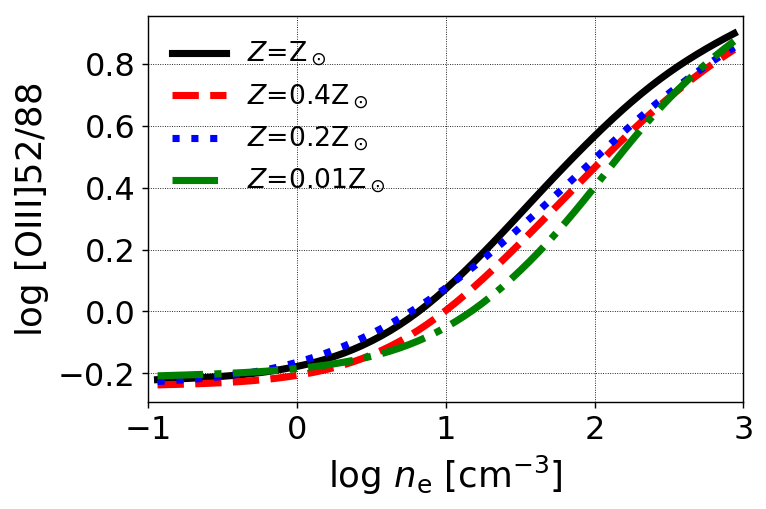}
  \caption{Line rations of [O$_{\rm ~III}]52$ with respect to [O$_{\rm ~III}]88$ computed with our model described in Section \ref{method}, as functions of electron density $n_{\rm e}$ for different metallicities $Z$. We here adopt our fiducial values of $t_{\rm HII}=1~{\rm Myr}$ and $M_{\rm cl}=10^5~{\rm M_\odot}$.}
  \label{LineRatio}
\end{figure}

In our model described above, the properties of a Str\"{o}mgren sphere are determined by an SED of a central star cluster and a local density $n_{\rm e}$, and the SED depends on $Z$, $t_{\rm HII}$ and $M_{\rm cl}$. Hence, if we fix $t_{\rm HII}$ and $M_{\rm cl}$, $U_{\rm in}$ becomes a function of $n_{\rm e}$ at a given $Z$. Fig. \ref{LineRatio} illustrates the line ratios of [O$_{\rm ~III}]52/88$ computed with our model, as functions of $n_{\rm e}$ for different $Z$. The line ratios monotonically increase with $n_{\rm e}$ and hardly depend on $Z$ (see also Fig. \ref{Udep} in Appendix \ref{app2}). Thus, [O$_{\rm ~III}]52/88$ is expected to be an observational tracer of a local density in a star-forming region. Combining it with the fact that intensities of the [O$_{\rm ~III}$] lines are nearly proportional to SFR \citep[see below and e.g.][]{dcl:14}, [O$_{\rm ~III}]52/88$ measured within an aperture can be used to estimate the averaged gas density weighted by SFR.

\subsection{Parameter calibration and relationship with SFR}
\label{calibration}
\begin{figure}
  \includegraphics[bb=0 0 1711 1198, width=\hsize]{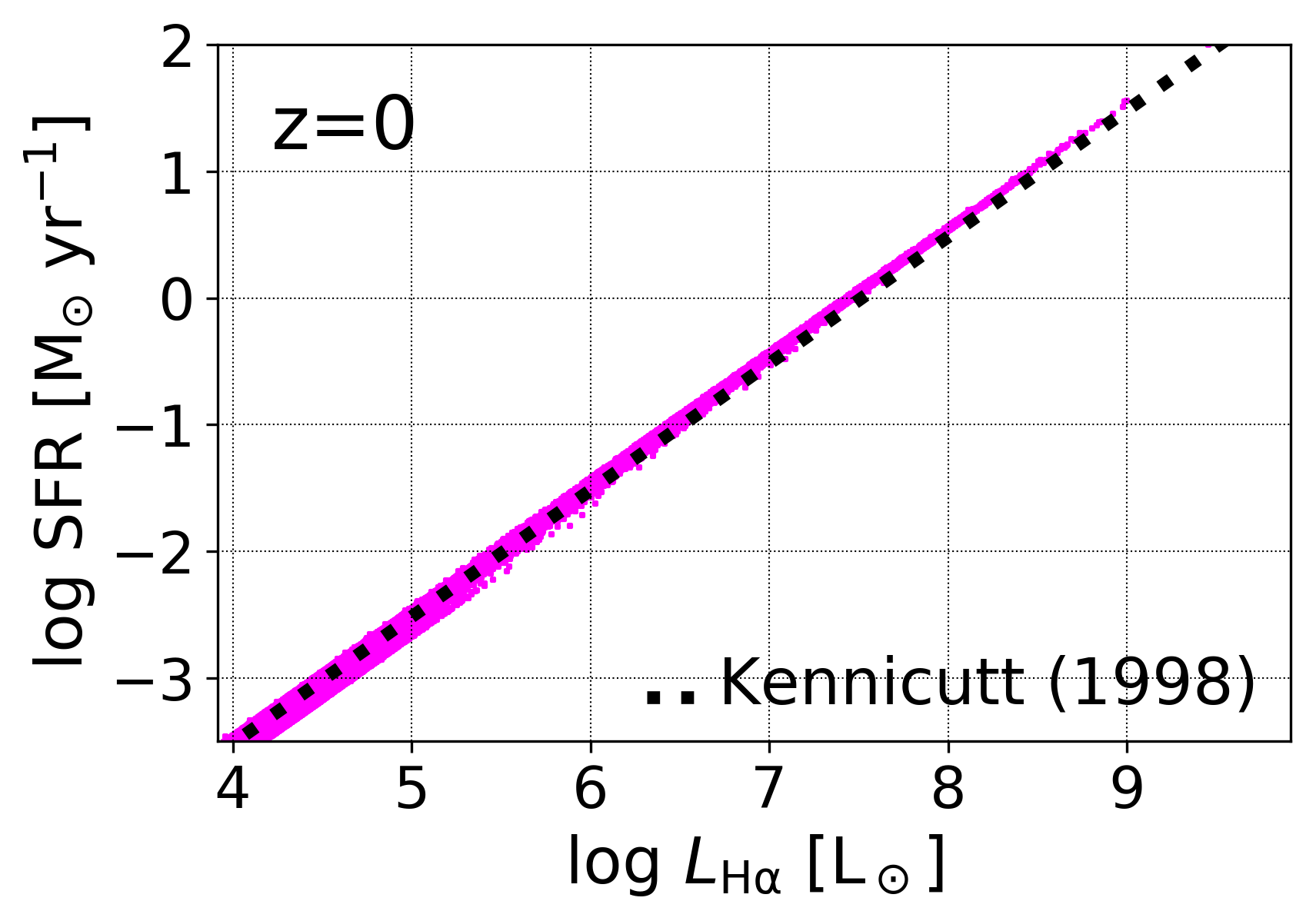}
  \caption{Correlation between H$\alpha$ luminosities and SFRs. The magenta dots indicate the galaxy-integrated values at redshift $z=0$ in IllustrisTNG. Although all galaxies in the simulation are plotted, their distribution highly concentrates along a linear relation. The black dotted line shows the observational result of \citet{k:98}.}
  \label{Kennicutt}
\end{figure}
Fig. \ref{Kennicutt} shows relationship between the galaxy-integrated SFRs and the modelled H$\alpha$ luminosities of all galaxies at redshift $z=0$ in IllustrisTNG. With the fiducial values of $t_{\rm HII}=1~{\rm Myr}$ and $M_{\rm cl}=10^5~{\rm M_\odot}$, our result is in agreement with the observed relation of \citet{k:98}. Our result for Illustris also shows the same consistency with the observations. We confirm that this result hardly depends on redshift or cluster mass in the ranges from $z=0$ to $6$ and from $M_{\rm cl}=10^3$ to $10^7~{\rm M_\odot}$.\footnote{When $M_{\rm cl}\lesssim10^2~{\rm M_\odot}$, we find that the total volumes of the Str\"{o}mgren spheres can exceed the CNM volumes in some cells. We consider that $M_{\rm cl}\gtrsim10^8~{\rm M_\odot}$ would be too large for a typical cluster mass.} The H$\alpha$ luminosities increase with $t_{\rm HII}$ and deviate from the observed relation if $t_{\rm HII}\gtrsim3~{\rm Myr}$. Accordingly, hereafter we argue our results with the fiducial values of $t_{\rm HII}=1~{\rm Myr}$ and $M_{\rm cl}=10^5~{\rm M_\odot}$.

\begin{figure}
  \includegraphics[bb=0 0 1705 1202, width=\hsize]{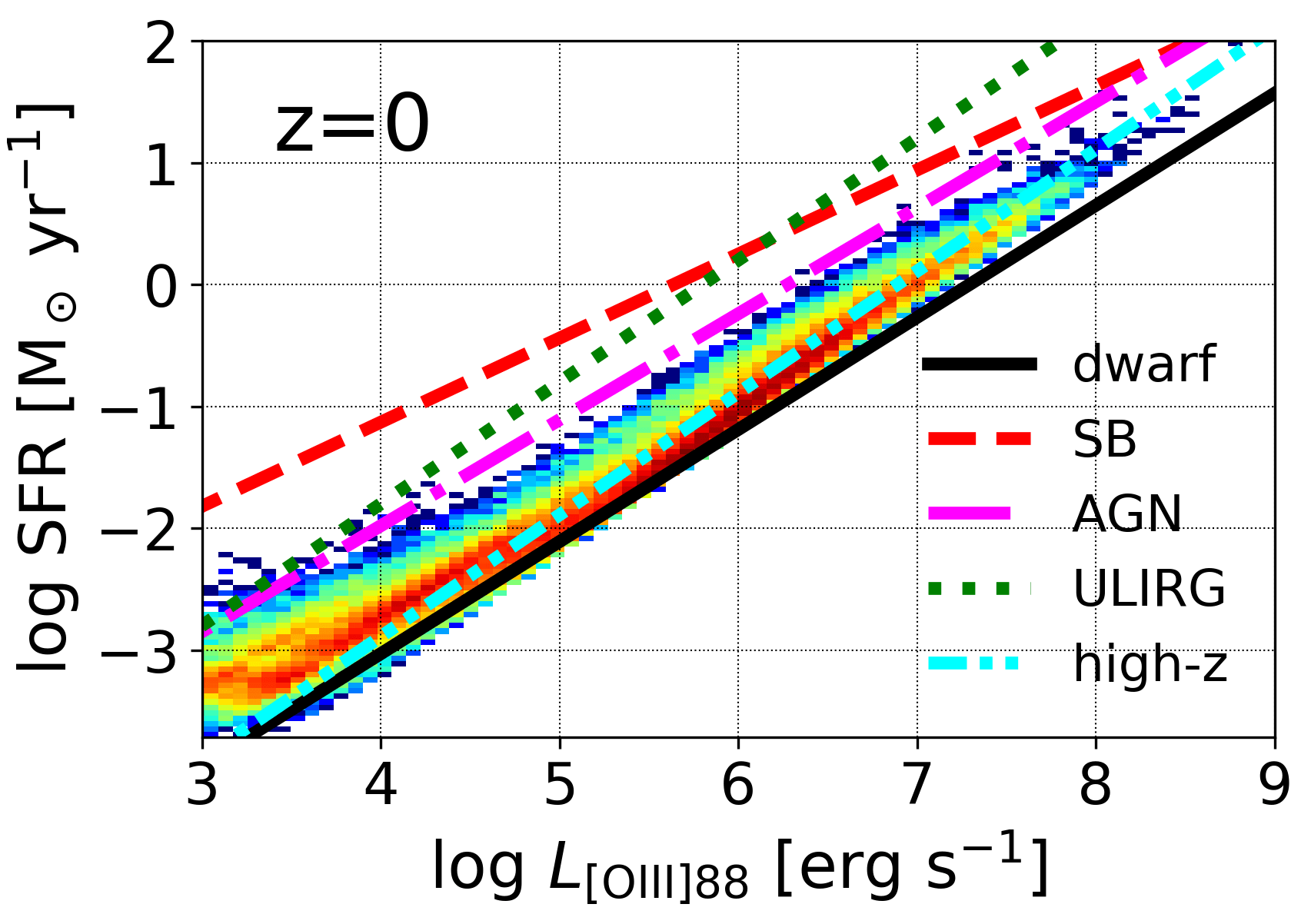}
  \caption{Correlation between luminosities of [O$_{\rm ~III}]88$ and SFRs of all galaxies in TNG at redshift $z=0$. The colour codes indicate the numbers of galaxies in each bin, increasing from blue to red in logarithmic scales. The straight lines show correlations observed by \citet{dcl:14} for the five types of galaxies: dwarfs, star burst (SB) galaxies, active galactic nuclei (AGNs), ultraluminous infrared galaxies (ULIRG) and high-redshift (high-$z$) galaxies.}
  \label{OIII_SFR}
\end{figure}
Fig. \ref{OIII_SFR} is the same as Fig. \ref{Kennicutt} but for luminosities of [O$_{\rm ~III}]88$ instead of H$\alpha$. Our result is in agreement with observations of \citet{dcl:14} for dwarf and high-redshift galaxies although the distribution of the simulated galaxies appears to be slightly below the observed relation for AGNs and also below the relations for star-burst and ultraluminous infrared galaxies. The correlation of the modelled [O$_{\rm ~III}]88$ and SFR little depends on $M_{\rm cl}$ or evolves with redshift $z$. We find no systematic differences of the correlation in Fig. \ref{OIII_SFR} between the star-forming and quiescent galaxies or between BHs in the high- and low-accretion modes. We find essentially the same result with Illustris.

\section{Results}
\label{Res}
We show our results for IllustrisTNG in Section \ref{ResTNG}. First, in Section \ref{SFMS}, we address characteristic features of the inside-out quenching processes caused by massive BHs in the simulation, i.e. depletion of central gas by the kinematic feedback. Next, in Section \ref{OIII_MBH_TNG}, we show how the modelled [O$_{\rm ~III}$] emission reflects the gas distribution affected by the BH quenching. In Section \ref{ResIll}, we follow the same course but show the results for Illustris.

\subsection{IllustrisTNG}
\label{ResTNG}
\subsubsection{Influence of the BH feedback in star-forming galaxies}
\label{SFMS}
\begin{figure}
  \includegraphics[bb=0 0 1223 1106, width=\hsize]{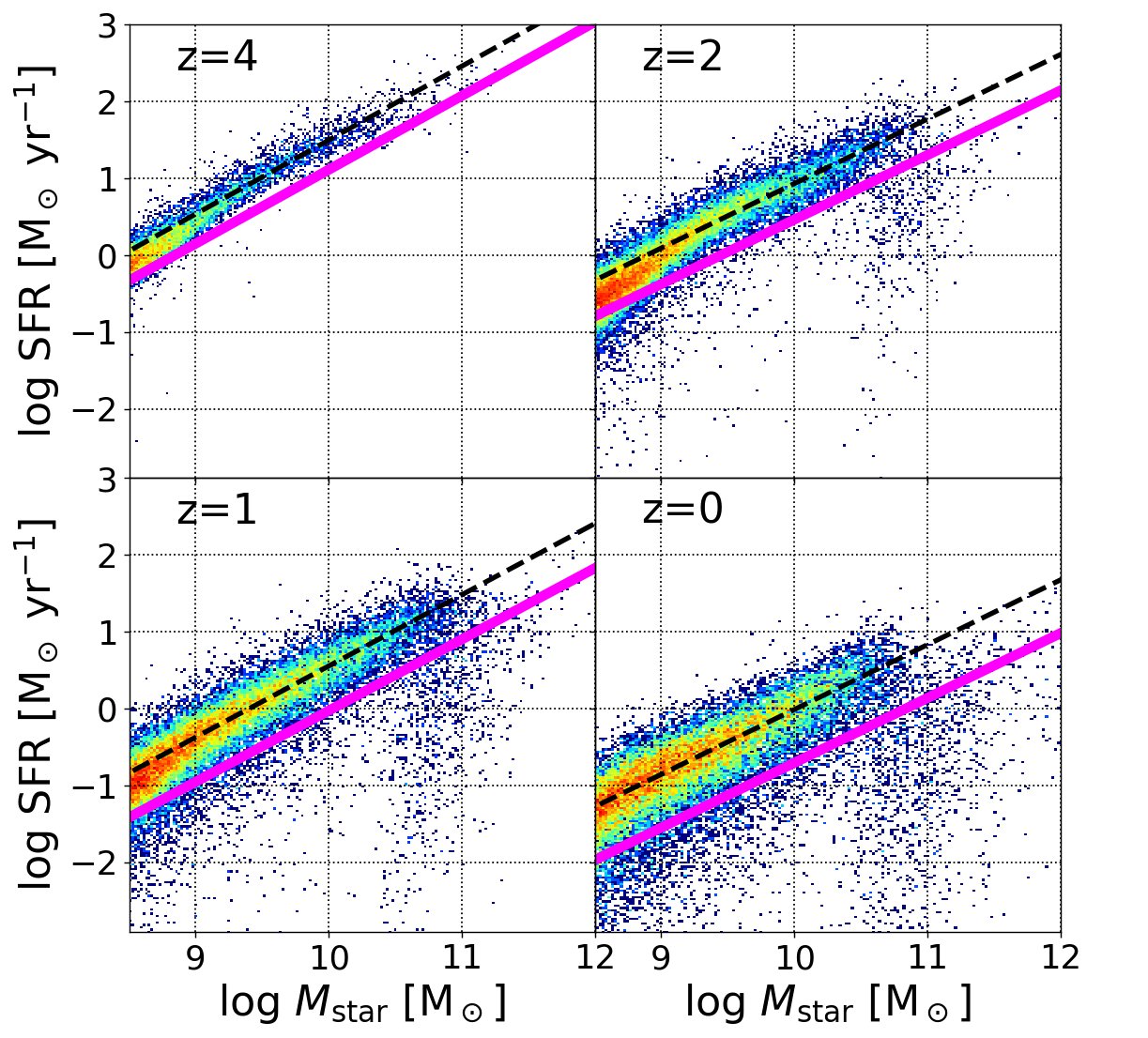}
  \caption{Distribution of the total SFRs and stellar masses of galaxies in TNG at redshifts $z=4$, $2$, $1$ and $0$. In each panel, the SFMS is indicated with the black dashed line. At a given $M_{\rm star}$, the boundary between star-forming and quenched galaxies (the magenta solid line) is defined to be an SFR at the offset $\Delta_{\rm MS}=-2.5\sigma_{\rm SFR}$ from the SFMS, where $\sigma_{\rm SFR}$ is the averaged dispersion of SFRs (see Appendix \ref{App1}). The colour codes indicate the numbers of galaxies in each bin, increasing from blue to red in logarithmic scales.}
  \label{SFMSTNG}
\end{figure}
To sample star-forming galaxies from the simulation data, we define a star-formation main sequence (SFMS: correlation between the total stellar masses $M_{\rm star}$ and SFRs) in each snapshot. We use a method proposed by \citet{dpn:19} to both simulations, describe the method in Appendix \ref{App1}. Fig. \ref{SFMSTNG} illustrates the distribution of $M_{\rm star}$ and SFRs of galaxies in IllustrisTNG, and the black dashed line delineates the SFMS defined at each redshift. The magenta solid line is the boundary between star-forming and quiescent galaxies, and this study samples those above the boundary as star-forming galaxies. 

\begin{figure}
  \includegraphics[bb=0 0 672 349, width=\hsize]{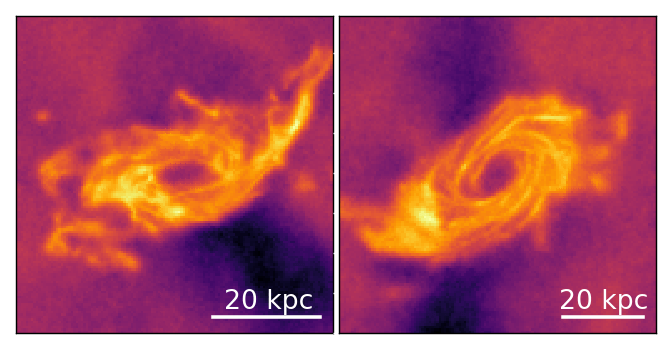}
  \caption{Projected maps of surface gas densities in two examples of star-forming galaxies in TNG at $z=0$. They have cavities in gas distribution around BHs. The orientations of the discs are not exactly face-on or edge-on. The left and right galaxies have $M_{\rm BH}=4.6$ and $2.4\times10^9~{\rm M_\odot}$, $M_{\rm star}=14.2$ and $9.2\times10^{11}~{\rm M_\odot}$ and SFRs of $100$ and $35~{\rm M_\odot~yr^{-1}}$, respectively.}
  \label{CObrightest}
\end{figure}
\begin{figure}
  \includegraphics[bb=0 0 1204 598, width=\hsize]{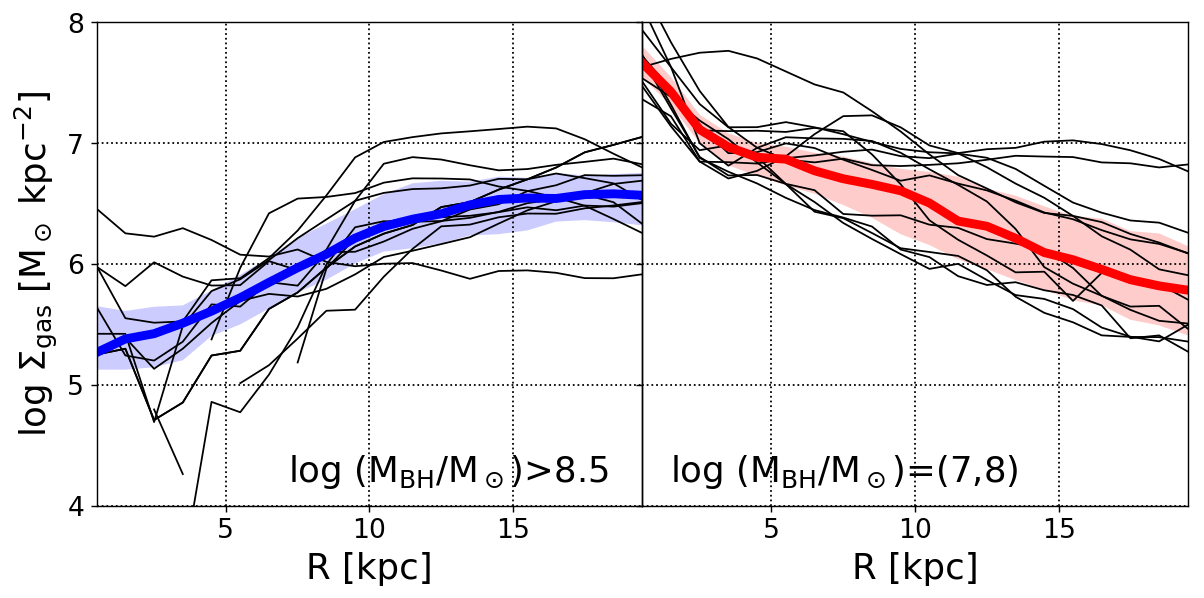}
  \caption{Radial profiles of face-on gas surface densities of the star-forming galaxies in IllustrisTNG at redshift $z=0$. The left and right panels show galaxies in the BH-mass ranges of $\log(M_{\rm BH}/{\rm M_\odot})>8.5$ and $7<\log(M_{\rm BH}/{\rm M_\odot})<8$. In each panel, the thick line delineates the stacked profile among 128 galaxies selected randomly in the BH-mass range, and the shaded region indicates the range of $\pm1\sigma$. The thin lines delineate profiles of 12 individual galaxies selected randomly. The left panel shows that the gas densities decrease towards the central BHs in the galaxies with the high BH masses.}
  \label{gasdensity}
\end{figure}
Fig. \ref{CObrightest} exemplifies two star-forming galaxies in IllustrisTNG at redshift $z=0$. Both have BHs more massive than $10^9~{\rm M_\odot}$ and high SFRs. With molecular gas modelling of \citet{iyy:20}, these galaxies are found to be the first and eighth brightest galaxies in CO(1-0) emission in TNG100-1. These galaxies have `cavities' in their gas distributions at the galactic centres \citep[see also][]{tbp:20}. Fig. \ref{gasdensity} indicates gas surface density distributions $\Sigma_{\rm gas}$, which are the face-on radial profiles stacked among 128 galaxies randomly selected in the ranges of $\log(M_{\rm BH}/{\rm M_\odot})>8.5$ (left panel) and $7<\log(M_{\rm BH}/{\rm M_\odot})<8$ (right panel). The galaxies with massive BHs show $\Sigma_{\rm gas}$ decreasing towards their galactic centres, whereas those with less massive BHs have nearly exponential profiles in which $\Sigma_{\rm gas}$ is generally the highest at the centre. \citet{nps:19} have also shown that massive galaxies with $M_{\rm star}\gtrsim10^{10.5}~{\rm M_\odot}$ indicate SFRs decreasing in their central regions in the TNG50 simulation \citep[see also][]{dpj:20}. Thus, central gas densities can be significantly affected by the feedback of massive BHs in star-forming galaxies in TNG. Such characteristic features may be consistent with observations of green-valley galaxies (see Section \ref{dis_expect}).

\begin{figure}
  \includegraphics[bb=0 0 1146 1096, width=\hsize]{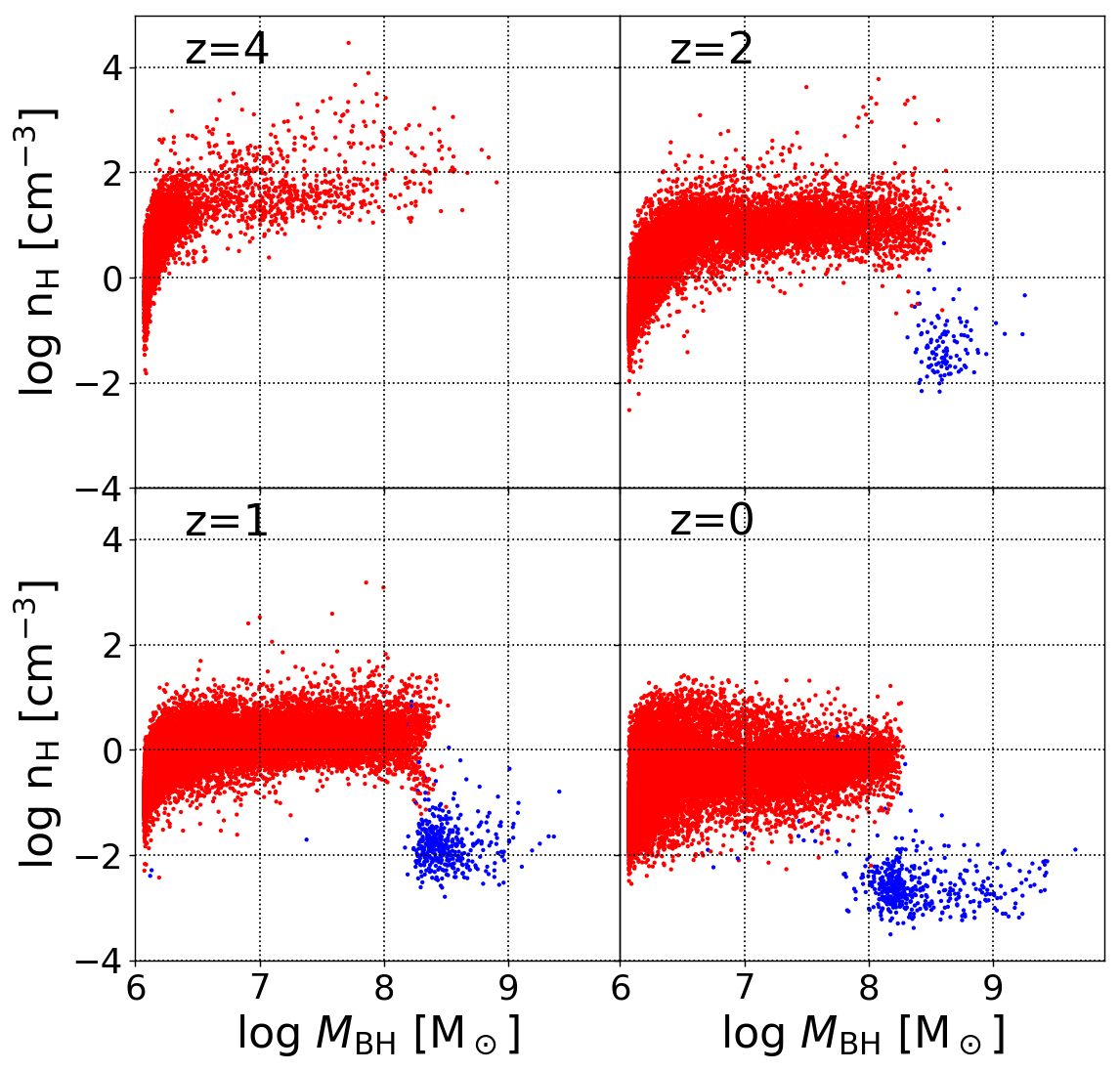}
  \caption{Gas densities $n_{\rm H}$ around BHs as functions of $M_{\rm BH}$ for the star-forming galaxies in IllustrisTNG at redshifts $z=4$, $2$, $1$ and $0$. The red and blue dots correspond to BHs whose feedback is in the high- and low-accretion modes. The ordinates $n_{\rm H}$ are cell densities $\rho_{\rm cell}$ averaged over the nearest $256\pm4$ cells around the BHs with a spline kernel, but in the units of number of hydrogen atoms per cubic centimetre.}
  \label{mbh-rho_tng}
\end{figure}
Fig. \ref{mbh-rho_tng} shows relationship between $M_{\rm BH}$ and gas densities around the BHs for the star-forming galaxies in IllustrisTNG. The densities around BHs abruptly decrease at $M_{\rm BH}\sim10^{8.5}~{\rm M_\odot}$, which is consistent with the presence of the cavities shown in Fig. \ref{CObrightest} and the left panel of Fig. \ref{gasdensity}. In Fig. \ref{mbh-rho_tng}, the red and blue dots correspond to galaxies whose BHs are in the high- and low-accretion modes, and most of the massive BHs in the low-accretion mode (blue dots) are surrounded by low-density gas. Hence, the depletion of the central gas is caused by the kinetic feedback of the low-accretion mode in TNG, rather than other mechanisms such as stellar feedback and consumption of gas by star formation. 

Previous studies using the IllustrisTNG simulation have also shown similar results. \citet{tbp:20} show the averaged gas densities within galaxies to significantly decrease in those with $M_{\rm BH}\gtrsim10^{8.2}~{\rm M_\odot}$ at $z=0$ since energies injected by the kinetic feedback of the low-accretion mode exceed binding energies in the galaxies. They propose, therefore, that galaxies whose BHs reach the critical $M_{\rm BH}$ are suddenly quenched and abruptly decrease their specific SFRs \citep[see also][]{wsp:18}. \citet{wsh:17} show that the critical BH mass at which the feedback modes are switched is $M_{\rm BH}\sim10^8$--$10^{8.5}~{\rm M_\odot}$, does not significantly vary with redshift in the case of the fiducial settings of IllustrisTNG\footnote{In TNG100-1, no BHs are in the low-accretion mode in redshifts $z\gtrsim4$. Therefore, the redshift-dependence of the critical $M_{\rm BH}$ is unclear above $z\sim4$.}. However, \citet{tbp:20} perform similar simulations but with different parameter settings and feedback models, demonstrate that the critical $M_{\rm BH}$ depends on the BH models and parameters therein. 

\subsubsection{The $[O_{\rm ~III}]$ line ratios and black hole masses}
\label{OIII_MBH_TNG}
Our result suggests that, if we observe the [O$_{\rm ~III}]$ lines from the simulated galaxies in TNG, the large difference of the central gas densities between galaxies whose $M_{\rm BH}$ are are higher and lower than the critical $M_{\rm BH}\sim10^{8.5}~{\rm M_\odot}$ can be inferred. To model the [O$_{\rm ~III}]$ line emission, we adopt our method described in Section \ref{modellinglines} to snapshots of TNG at redshifts $z=0$, $1$, $2$ and $4$.

\begin{figure*}
  \begin{center}
    \begin{tabular}{c}
    
      \begin{minipage}{0.5\hsize}
        \begin{center}
          \includegraphics[bb=0 0 1170 1092, width=\hsize]{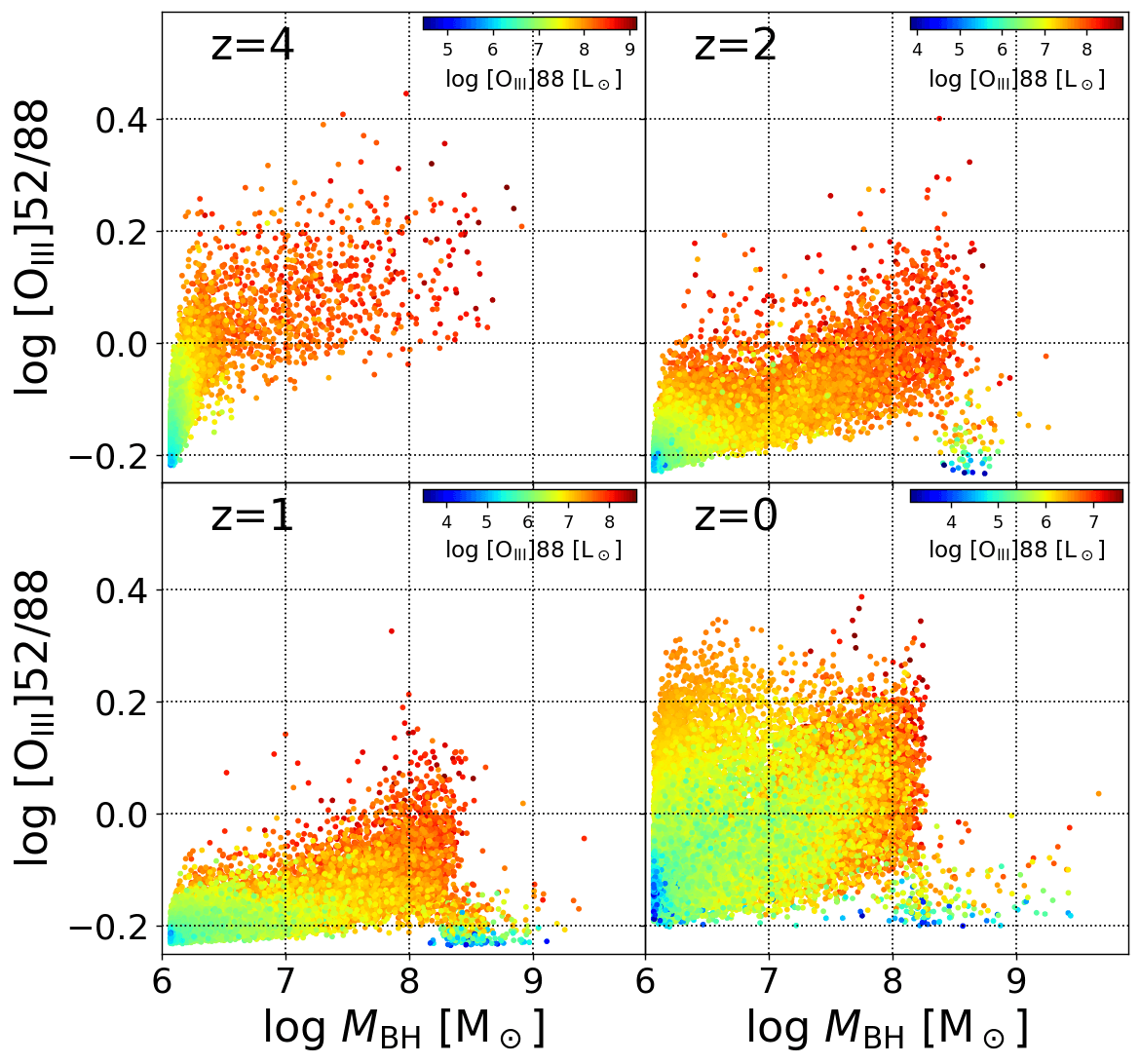}
        \end{center}
      \end{minipage}
      
      \begin{minipage}{0.5\hsize}
        \begin{center}
          \includegraphics[bb=0 0 1170 1092, width=\hsize]{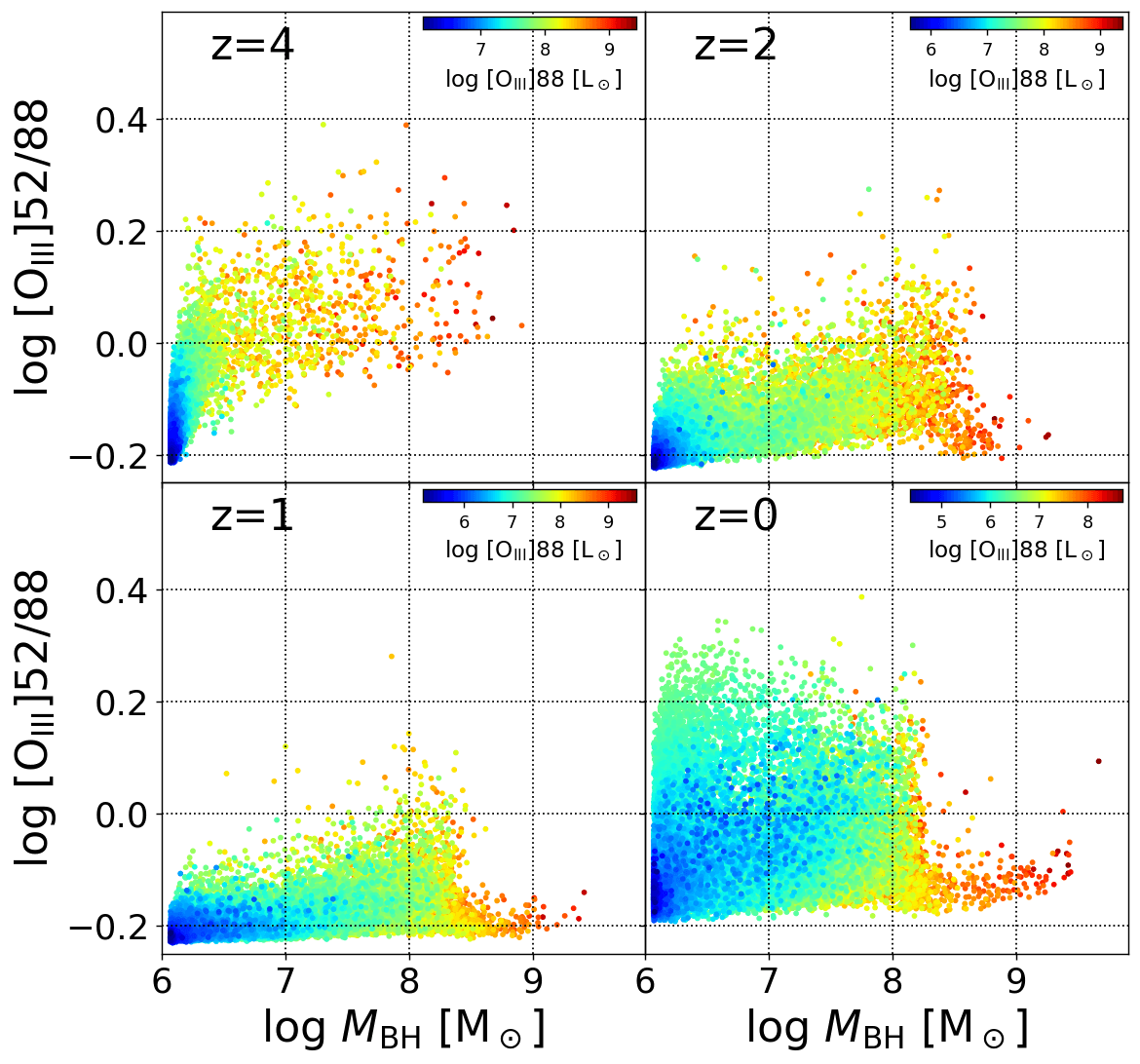}
        \end{center}
      \end{minipage}
    
    \end{tabular}
    \caption{Distribution of the line ratios of [O$_{\rm ~III}$]52 to [O$_{\rm ~III}$]88 and BH masses in the star-forming galaxies at redshifts $z=4$, $2$, $1$ and $0$ in IllustrisTNG. The line ratios are measured within a two-dimensional aperture of $R=3~{\rm kpc}$ centred on the BHs (left) and those covering the entire galaxies (right). The colours of the plotted dots indicate the luminosities of [O$_{\rm ~III}]88$ integrated within the apertures. Note that the colour scales are different between the panels.}
    \label{OIII_TNG}
  \end{center}
\end{figure*}
Fig. \ref{OIII_TNG} shows our results of modelled line ratios of [O$_{\rm ~III}]52/88$ as functions of $M_{\rm BH}$ for the star-forming galaxies in IllustrisTNG. In the left set of four panels, the [O$_{\rm ~III}]52/88$ ratios are measured within two-dimensional apertures of $R=3~{\rm kpc}$ centred on their BHs, where line-of-sight orientations of the galaxies are at random. The colours of the dots indicate unattenuated luminosities of [O$_{\rm ~III}]88$ integrated within the apertures. In TNG, no BHs are in the low-accretion mode at high redshifts $z\gtrsim4$ (see Fig. \ref{mbh-rho_tng}). At redshift $z=4$, [O$_{\rm ~III}]52/88$ appears to increase with $M_{\rm BH}$ although the scatter is large, which means that more massive BHs are embedded within gas with higher densities. The higher gas densities around more massive BHs are also reflected in stronger emission of [O$_{\rm ~III}]88$ which indicates higher SFRs within the apertures. In the following snapshots at $z=2$, $1$ and $0$, similar trends are only seen for galaxies hosting less massive BHs with $M_{\rm BH}\lesssim10^{8.5}~{\rm M_\odot}$. It is noteworthy that the line ratios of [O$_{\rm ~III}]52/88$ suddenly decrease at $M_{\rm BH}\sim10^{8.5}~{\rm M_\odot}$, and the [O$_{\rm ~III}]88$ luminosities within the 3-kpc aperture are weak in the galaxies with $M_{\rm BH}\gtrsim10^{8.5}~{\rm M_\odot}$. In comparing Figs. \ref{OIII_TNG} with \ref{mbh-rho_tng}, it is obvious that the drop of [O$_{\rm ~III}]52/88$ at $M_{\rm BH}\sim10^{8.5}~{\rm M_\odot}$ coincides with the transition from the high- to low-accretion modes of the BH feedback in TNG. Namely, the drop of [O$_{\rm ~III}]52/88$ reflects the depletion of central gas densities by the kinematic feedback of massive BHs: the early phase of the inside-out quenching. 

The right set of panels of Fig. \ref{OIII_TNG} shows [O$_{\rm ~III}]52/88$ measured within apertures covering the whole galaxies, and the ratios are not significantly different from those measured within the 3-kpc apertures. This is because, for galaxies with $M_{\rm BH}\lesssim10^{8.5}~{\rm M_\odot}$, their SFR distributions are compact, and their central regions inside $R=3~{\rm kpc}$ are therefore dominant in their [O$_{\rm ~III}$] emission. For galaxies with $M_{\rm BH}\gtrsim10^{8.5}~{\rm M_\odot}$, on the other hand, their central gas densities are significantly lowered by the kinematic feedback, and therefore mainly their outer regions contribute to the total SFRs and [O$_{\rm ~III}$] emission. However, such outer gas is not dense enough to raise the integrated [O$_{\rm ~III}]52/88$. It can be deduced from the result that the galaxy-integrated [O$_{\rm ~III}]88$ is high in the galaxies with massive BHs of $M_{\rm BH}\gtrsim1
0^{8.5}~{\rm M_\odot}$, unlike in the case of the 3-kpc apertures. Thus, from the above result, we predict that the abrupt decrease of [O$_{\rm ~III}]52/88$ is observed at the critical BH mass of $M_{\rm BH}\sim10^{8.5}~{\rm M_\odot}$ if the BH model implemented in IllustrisTNG is accurate. In addition, the similarity between [O$_{\rm ~III}]52/88$ measured within the 3-kpc and galaxy-integrated apertures means that observations for the line ratios do not need to resolve the central regions of galaxies (although see below).
 
Although BH masses have been measured in a substantial number of galaxies, determinations of $M_{\rm BH}$ basically require different observations independent from measurements for the [O$_{\rm ~III}$] lines. It has been known that $M_{\rm BH}$ correlates with stellar velocity dispersions (VDs) of spheroidal components of galaxies: the $M$--$\sigma$ relation \citep[e.g.][and references therein]{fm:20,gbb:00,ms:12}. This implies that more massive BHs generally reside in more massive galaxies although there are scatters to some extent in the correlation. It may be expected that the large masses of galaxies can enlarge not only stellar VDs but also gaseous ones. We compute the second-moment of [O$_{\rm ~III}]88$ emission as
\begin{equation}
\sigma_{\rm [O_{III}]88}^2=\frac{\int l_{\rm cell}\left(v_{\rm los}-\overline{v_{\rm los}}\right)^2 \mathrm{d}S}{\int l_{\rm cell}\mathrm{d}S},
\label{SecMom}
\end{equation}
where $l_{\rm cell}$ and $v_{\rm los}$ are [O$_{\rm ~III}]88$ luminosity and line-of-sight velocity of a single gas cell, and the integrals in the denominator and numerator cumulate gas cells within the observational apertures. The mean velocity $\overline{v_{\rm los}}$ is measured within the aperture: $\overline{v_{\rm los}}=\int l_{\rm cell}v_{\rm los}\mathrm{d}S/\int l_{\rm cell}\mathrm{d}S$. We consider that equation (\ref{SecMom}) corresponds to a VD weighted by [O$_{\rm ~III}]88$ luminosity. 

\begin{figure*}
  \begin{center}
    \begin{tabular}{c}
    
      \begin{minipage}{0.5\hsize}
        \begin{center}
          \includegraphics[bb=0 0 1149 1092, width=\hsize]{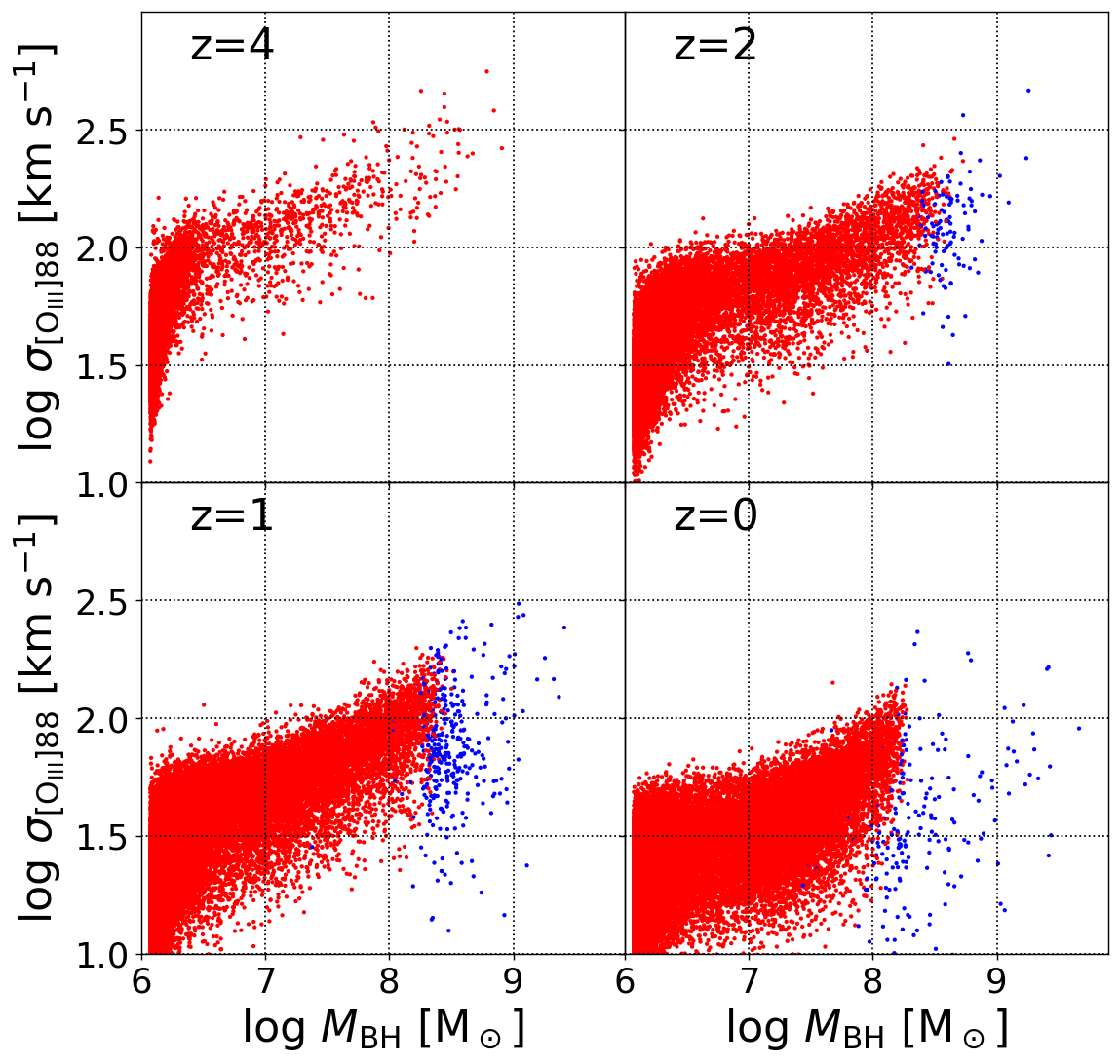}
        \end{center}
      \end{minipage}
      
      \begin{minipage}{0.5\hsize}
        \begin{center}
          \includegraphics[bb=0 0 1149 1092, width=\hsize]{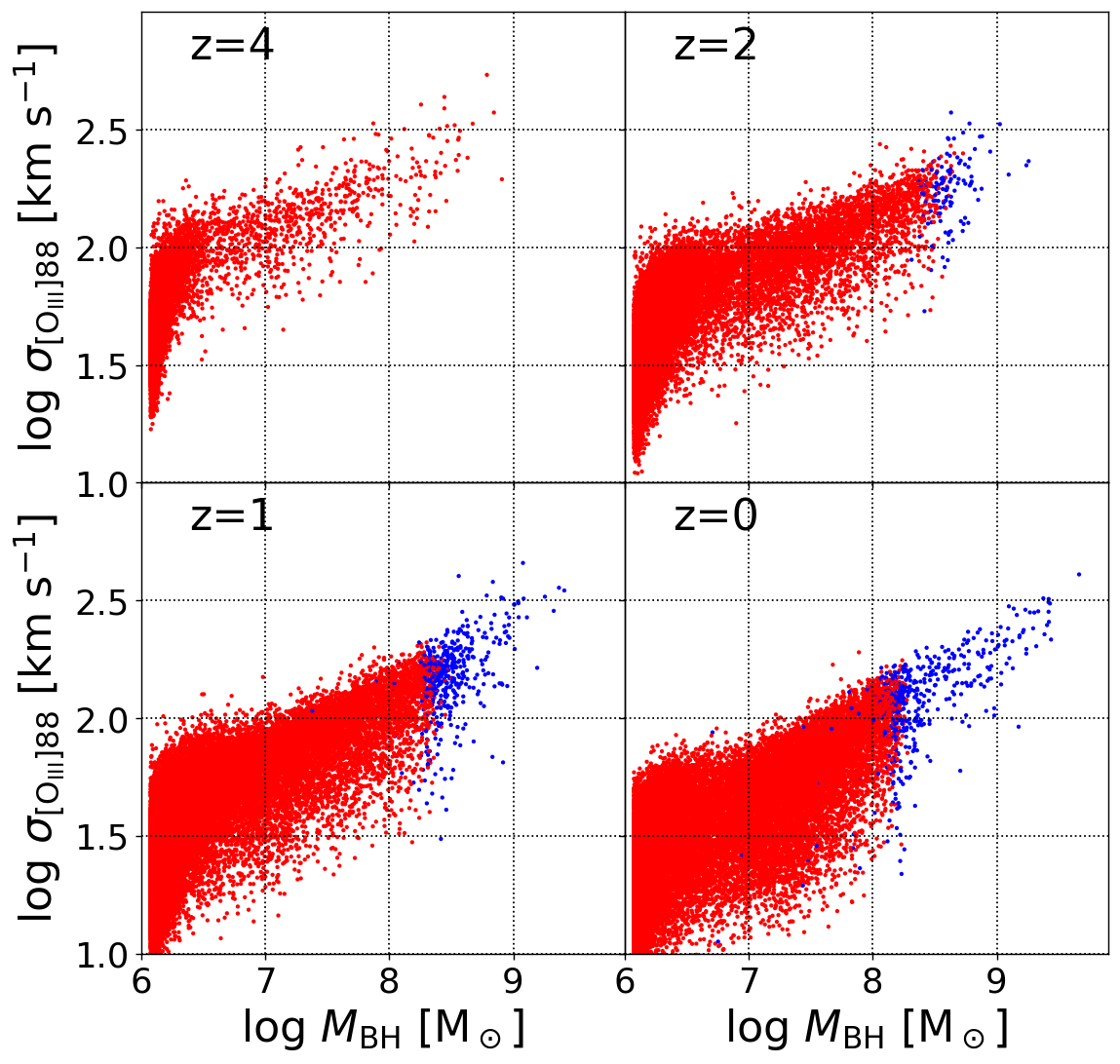}
        \end{center}
      \end{minipage}
    
    \end{tabular}
    \caption{Correlations between $M_{\rm BH}$ and the second-moments of [O$_{\rm ~III}]88$ lines at redshifts $z=4$, $2$, $1$ and $0$ in IllustrisTNG. In the left and right sets of panels, the 3-kpc and galaxy-integrated apertures are applied to compute $\sigma_{\rm [O_{III}]88}$. As in the Fig. \ref{mbh-rho_tng}, the red and blue bots indicate galaxies whose BH feedback is in the high- and low accretion modes.}
    \label{BH_VD_TNG}
  \end{center}
\end{figure*}
Fig. \ref{BH_VD_TNG} shows the distribution of $\sigma_{\rm [O_{III}]88}$ and $M_{\rm BH}$ in IllustrisTNG. In the case of the 3-kpc apertures (the left set of panels), although the galaxies in the high-accretion mode (the red dots) indicate the correlation between their $\sigma_{\rm [O_{III}]88}$ and $M_{\rm BH}$, those in the low-accretion mode (the blue dots) have significantly lower $\sigma_{\rm [O_{III}]88}$ than the extrapolation of the correlation. In the case of the galaxy-integrated apertures (the right set of panels), on the other hand, $\sigma_{\rm [O_{III}]88}$ appears to correlate with $M_{\rm BH}$ regardless of the BH feedback modes. We infer that BHs in the kinematic mode blow out gas around the galactic centres to outer regions and would significantly reduce dynamical masses inside the 3-kpc apertures. Hence, virialised VDs inside $3~{\rm kpc}$ result in low values at a given $M_{\rm BH}$ above the critical mass $M_{\rm BH}\sim10^{8.5}~{\rm M_\odot}$. The galaxy-integrated apertures capture all gas belonging to the galaxies although $\sigma_{\rm [O_{III}]88}$ would be biased to high-SFR regions. Since BH masses are thought to correlate with the total masses of the galaxies, the galaxy-integrated $\sigma_{\rm [O_{III}]88}$ reflecting the total mass still holds the correlation. From this result, we expect that $M_{\rm BH}$ can be replaced with $\sigma_{\rm [O_{III}]88}$ if the galaxy-integrated apertures are applied. 

\begin{figure*}
  \begin{center}
    \begin{tabular}{c}
    
      \begin{minipage}{0.5\hsize}
        \begin{center}
          \includegraphics[bb=0 0 1170 1101, width=\hsize]{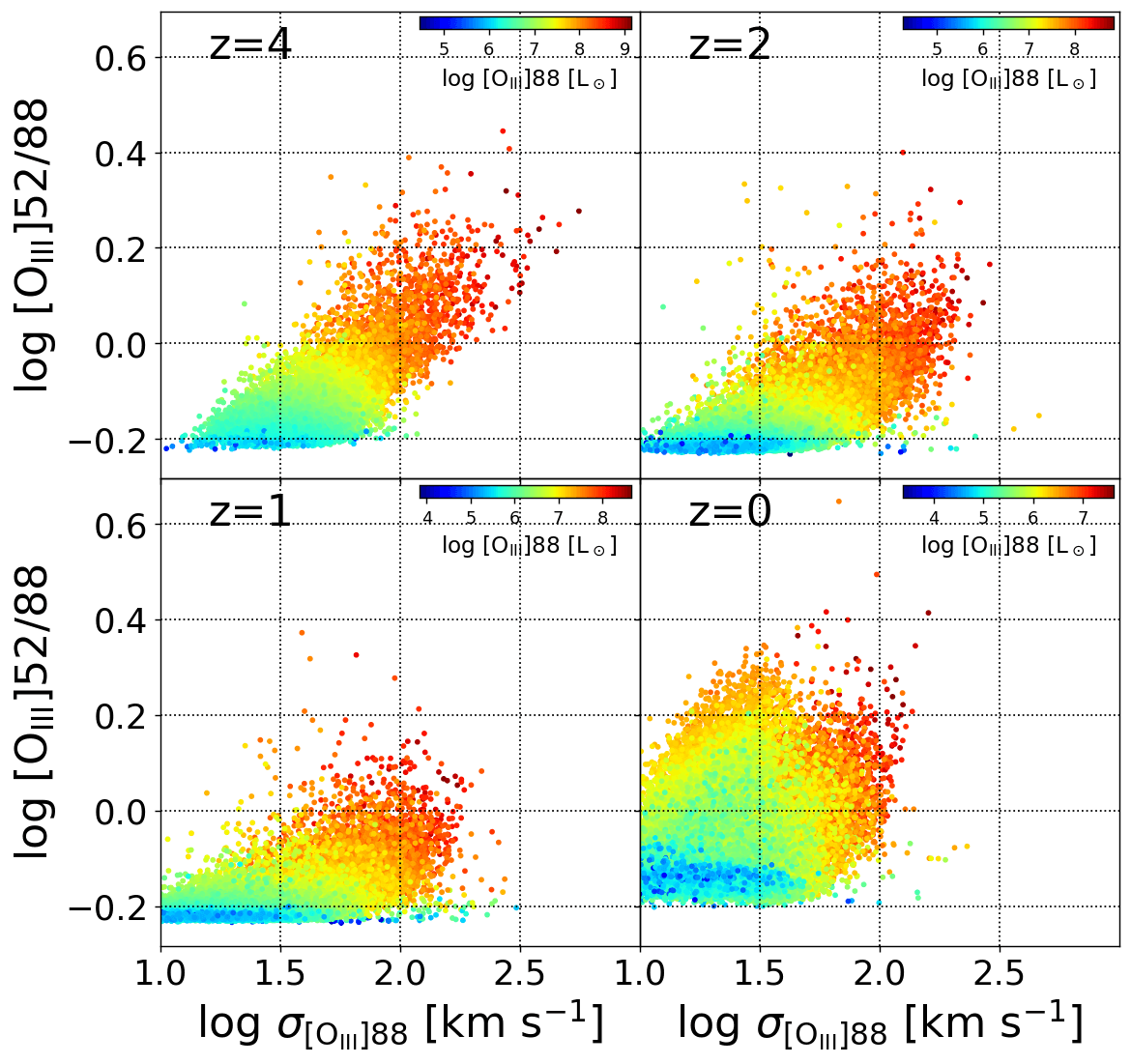}
        \end{center}
      \end{minipage}
      
      \begin{minipage}{0.5\hsize}
        \begin{center}
          \includegraphics[bb=0 0 1170 1101, width=\hsize]{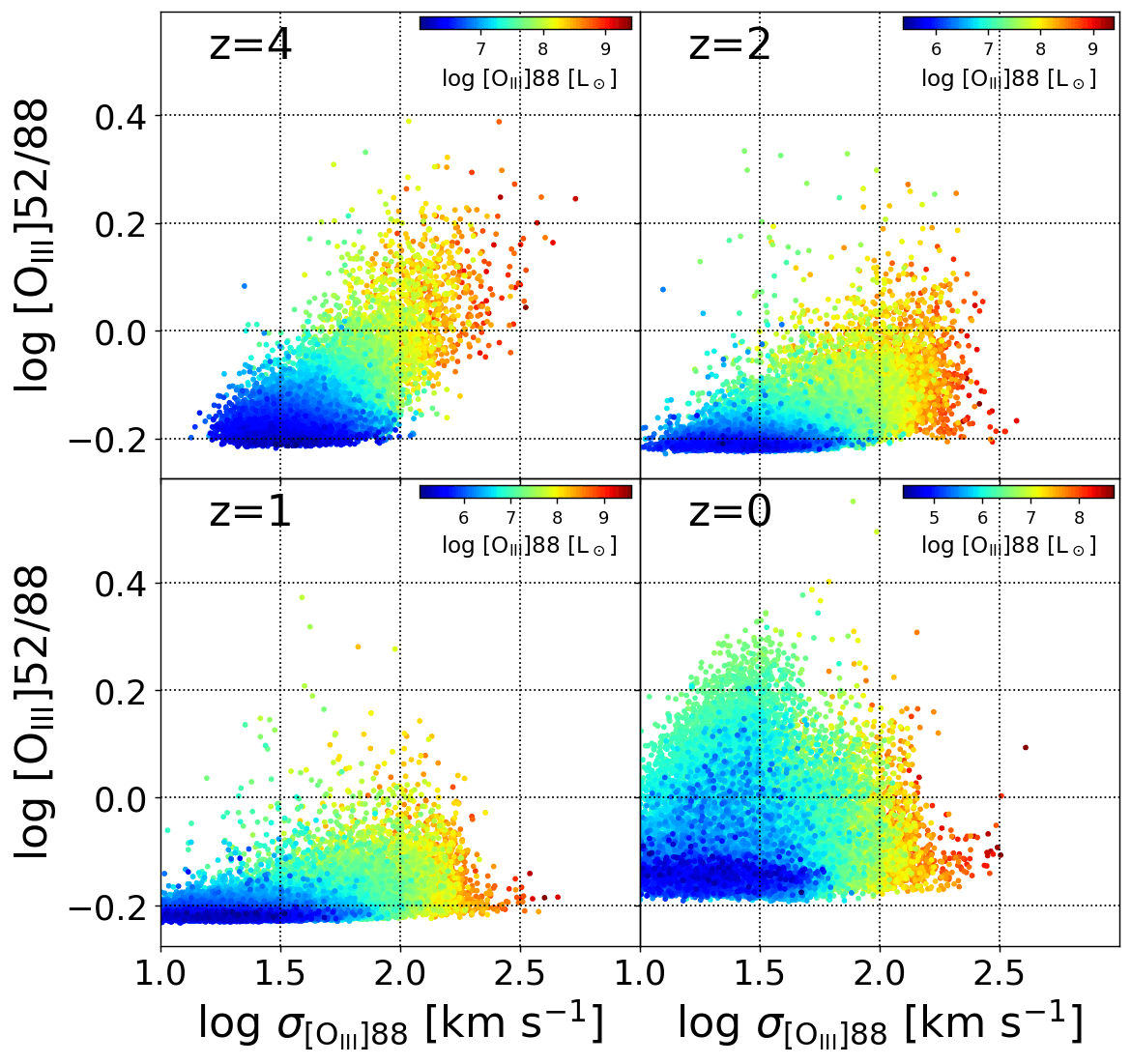}
        \end{center}
      \end{minipage}
    
    \end{tabular}
    \caption{Same as Fig. \ref{OIII_TNG} but with $\sigma_{\rm [O_{III}]88}$ instead of $M_{\rm BH}$ on the abscissas.}
    \label{VD_MBH_TNG}
  \end{center}
\end{figure*}
Fig. \ref{VD_MBH_TNG} shows the line ratios of [O$_{\rm ~III}]52/88$ as functions of $\sigma_{\rm [O_{III}]88}$ for IllustrisTNG. As we expect above, since $\sigma_{\rm [O_{III}]88}$ does not correlate with $M_{\rm BH}$ above the critical BH mass in the case of the 3-kpc apertures, the drop of [O$_{\rm ~III}]52/88$ is not clearly seen in the left set of panels. On the other hand, in the case of the galaxy-integrated apertures, the abrupt decrease of [O$_{\rm ~III}]52/88$ is still apparent at $\sigma_{\rm [O_{III}]88}\sim200~{\rm km~s^{-1}}$ although it is less clear than in Fig. \ref{OIII_TNG}. Thus, if we use observational apertures large enough to cover nearly the entire regions of galaxies, we may be able to use the second-moment of [O$_{\rm ~III}]88$ instead of $M_{\rm BH}$ in searching for the drop of [O$_{\rm ~III}]52/88$. These two quantities can be obtained from the same spectroscopic observations.

\subsection{Illustris}
\label{ResIll}
As we mention in Section \ref{BHmodel}, the Illustris simulation is largely different from TNG in the low-accretion mode of the BH feedback. Instead of the kinematic feedback in TNG, Illustris adopts the thermal feedback designed to reproduce AGN bubbles to the BHs whose Eddington ratios are below the constant criterion $\chi=0.05$ (see Section \ref{BHmodel}). Therefore, the expected line ratios can differ between the two simulations, and their comparison may give us clues to constrain BH models and their parameters.

\begin{figure}
  \includegraphics[bb=0 0 1146 1096, width=\hsize]{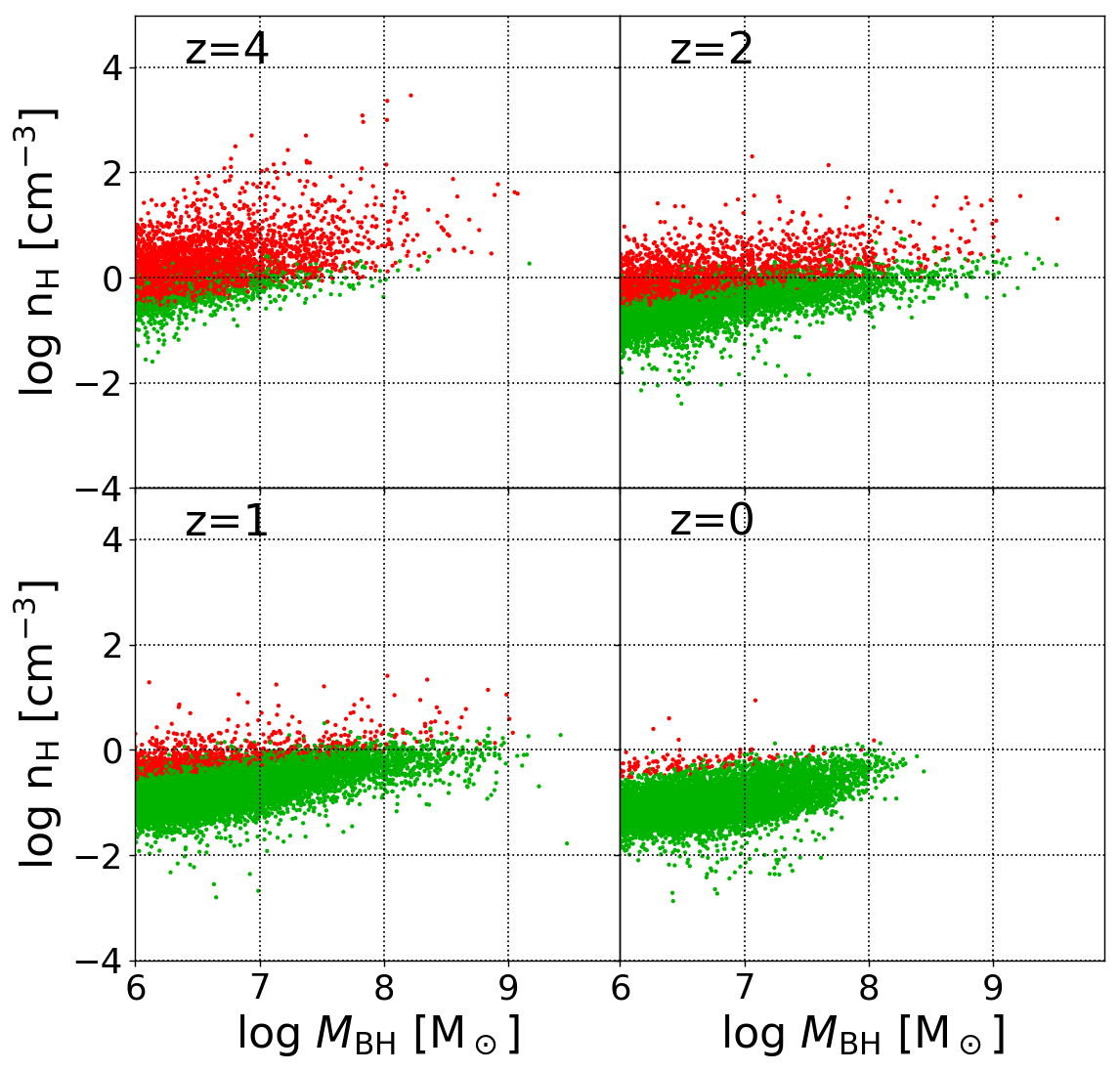}
  \caption{Same as Fig. \ref{mbh-rho_tng} but for the star-forming galaxies at $z=0$ in Illustris. The red and green dots correspond to BHs whose feedback is in the high- and low-accretion modes, respectively. Note that the criterion $\chi$ to switch the feedback modes and the implementation of the low-accretion mode are largely different from those of IllustrisTNG (see Section \ref{BHmodel}).}
  \label{mbh-rho_ill}
\end{figure}
\begin{figure}
  \includegraphics[bb=0 0 1204 598, width=\hsize]{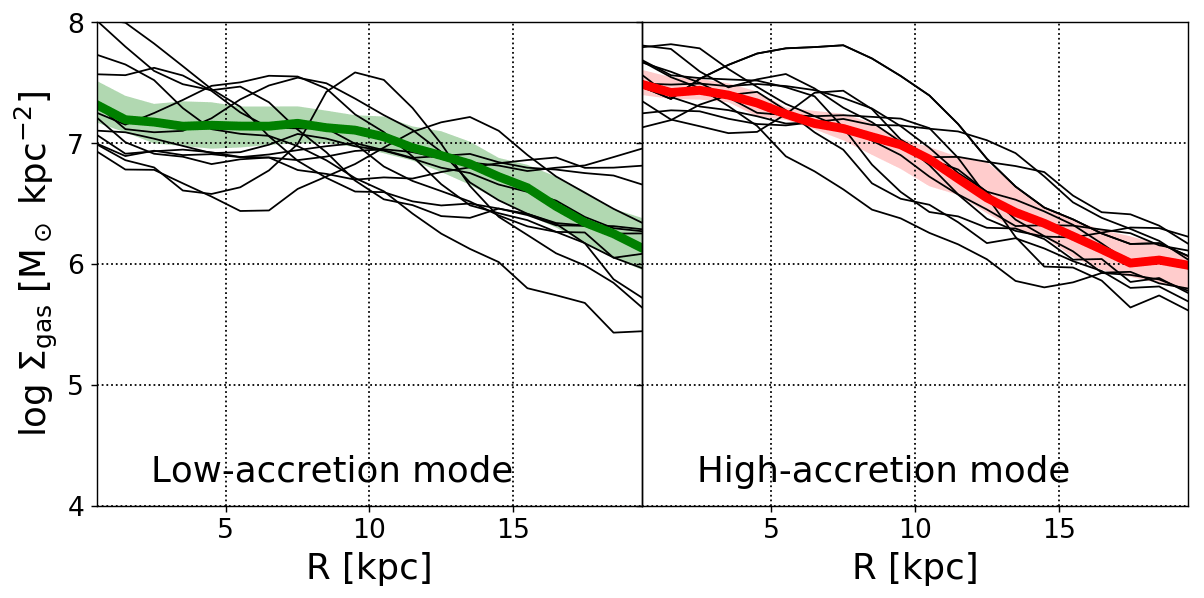}
  \caption{Radial profiles of face-on gas surface densities of the star-forming galaxies with $M_{\rm BH}>10^7~{\rm M_\odot}$ in Illustris at redshift $z=0$. The left and right panels show galaxies whose BHs are in the low- and high-accretion modes. Note that there are only a small number of galaxies in the high-accretion mode at $z=0$. Although the thick line in the left panel delineates the stacked profile among 128 galaxies selected randomly, that in the right panel samples only 54 galaxies. The shaded region indicates the range of $\pm1\sigma$. The thin lines delineate individual profiles of 12 galaxies selected randomly.}
  \label{gasdensity_Ill}
\end{figure}
As is done for TNG in Section \ref{SFMS}, we extract star-forming galaxies distributing along the SFMS in each snapshot and show the SFMSs in Appendix \ref{App1}. For the star-forming galaxies, Fig. \ref{mbh-rho_ill} shows the relationships between $M_{\rm BH}$ and gas densities around the BHs. In contrast with TNG (Fig. \ref{mbh-rho_tng}), the gas densities around BHs little change and weakly increase with $M_{\rm BH}$ in Illustris, and the transition of BH feedback between the high- and low-accretion modes does not appear to significantly affect the gas densities. In addition, there is not a characteristic value of $M_{\rm BH}$ to switch the feedback modes in Illustris. Unlike in TNG, no galaxies have BHs with $M_{\rm BH}\gtrsim10^{8.5}~{\rm M_\odot}$ at redshift $z=0$ in Illustris. This is because galaxies hosting such massive BHs are quenched after $z=1$ and drop off from the SFMS (see Fig. \ref{SFMSILL}). They are no longer classified as star-forming galaxies and excluded from our analysis. This behaviour is also due to the different feedback model for the low-accretion mode in Illustris. 

Fig. \ref{gasdensity_Ill} shows face-on radial profiles of $\Sigma_{\rm gas}$ of the star-forming galaxies with $M_{\rm BH}>10^7~{\rm M_\odot}$ at $z=0$, where the left and right panels illustrate those of galaxies in the low- and high-accretion modes. The stacked profiles are similar between galaxies in the low- and high-accretion modes although those in the low-accretion mode have slightly lower $\Sigma_{\rm gas}$ in their central regions. Note that the sample selection in Fig. \ref{gasdensity_Ill} is different from that in Fig. \ref{gasdensity}. However, these selections are essentially consistent since all BHs with $\log(M_{\rm BH}/{\rm M_\odot})>8.5$ and few BHs with $7<\log(M_{\rm BH}/{\rm M_\odot})<8$ are in the low-accretion mode in Fig. \ref{gasdensity}. The influence of the BH feedback on the central gas densities is thus largely different between TNG and Illustris because of their different modellings for the feedback, especially the low-accretion modes and the switching criteria $\chi$. As shown above, the low-accretion mode in the Illustris simulation is not influential on gas distribution.

\begin{figure*}
  \begin{center}
    \begin{tabular}{c}
    
      \begin{minipage}{0.5\hsize}
        \begin{center}
          \includegraphics[bb=0 0 1170 1092, width=\hsize]{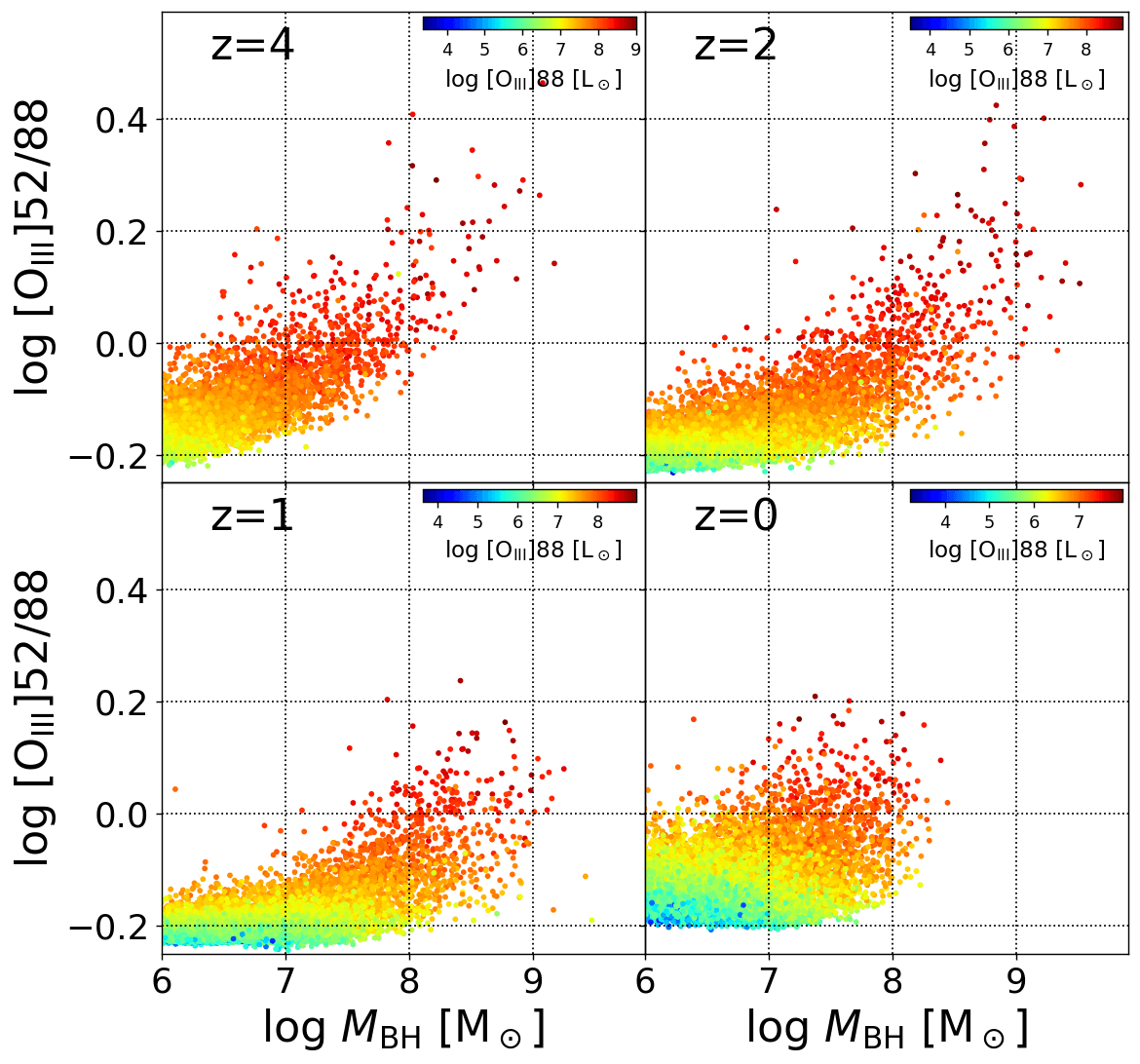}
        \end{center}
      \end{minipage}
      
      \begin{minipage}{0.5\hsize}
        \begin{center}
          \includegraphics[bb=0 0 1170 1092, width=\hsize]{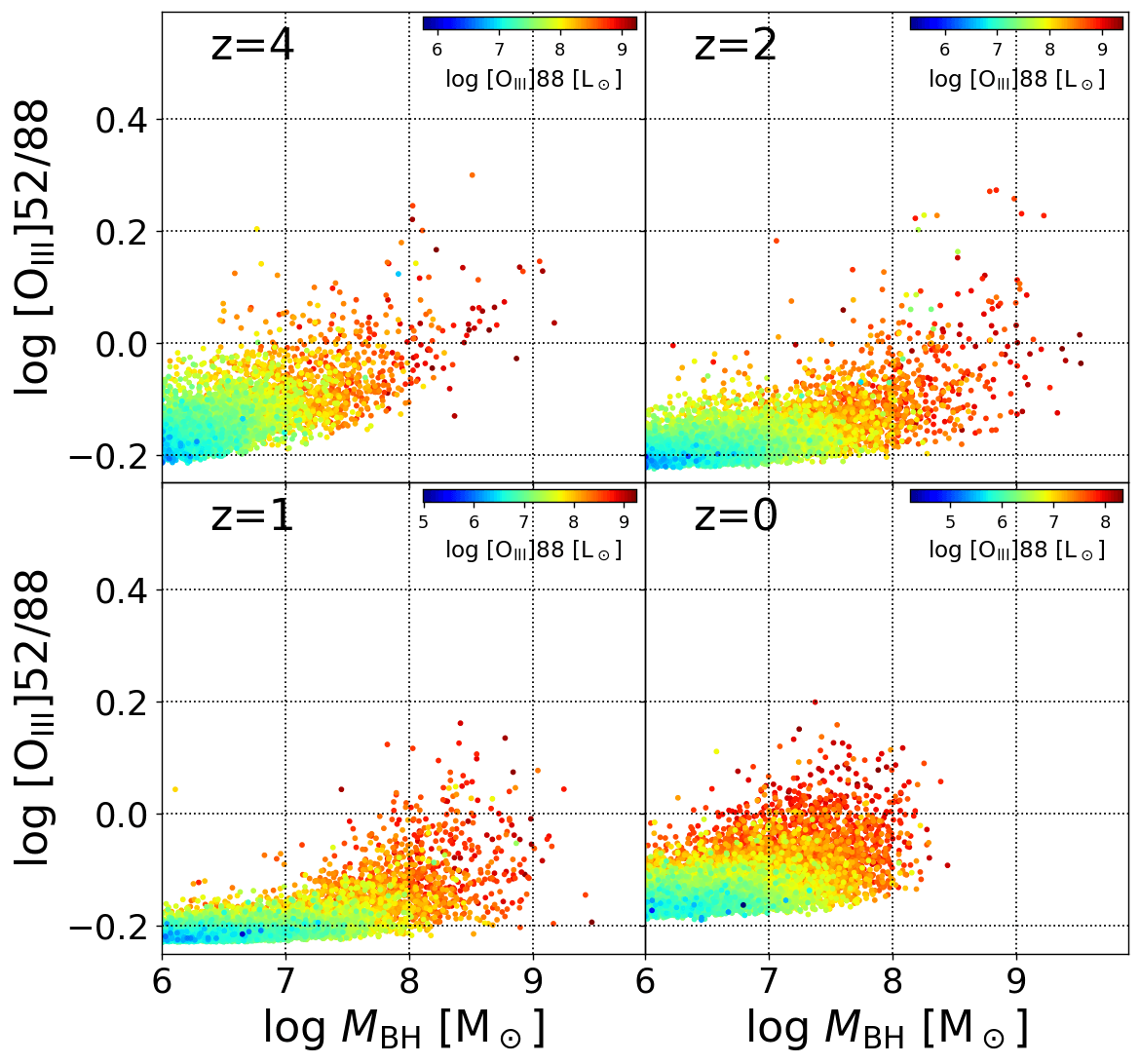}
        \end{center}
      \end{minipage}
    
    \end{tabular}
    \caption{Same as Fig. \ref{OIII_TNG} but for the Illustris simulation.}
    \label{OIII_Ill}
  \end{center}
\end{figure*}
Fig. \ref{OIII_Ill} shows the relationship between $M_{\rm BH}$ and [O$_{\rm ~III}]52/88$ for the star-forming galaxies in Illustris. As expected from Figs. \ref{mbh-rho_ill} and \ref{gasdensity_Ill}, there is no clear drop of [O$_{\rm ~III}]52/88$ at any $M_{\rm BH}$, and the line ratios continuously increase with $M_{\rm BH}$ although the scatters are large as well as in TNG. The galaxy-integrated values (right) of [O$_{\rm ~III}]52/88$ are not significantly different from those measured within the 3-kpc apertures (left). Thus, we predict that abrupt decrease of [O$_{\rm ~III}]52/88$ is not observed at any $M_{\rm BH}$ if the BH model of Illustris is accurate. Accordingly, the difference of the BH models in the simulations, especially the low-accretion modes of the feedback, is expected to significantly impact the line ratios of [O$_{\rm ~III}]52/88$ measured in star-forming galaxies hosting massive BHs. From these results, we propose that observations for [O$_{\rm ~III}]52/88$ can constrain the BH models implemented in simulations. As in the case of TNG (Fig. \ref{OIII_TNG}), the integrated [O$_{\rm ~III}]52/88$ does not significantly depend on aperture size. The reason can be deduced from Fig. \ref{gasdensity_Ill}; the averaged profiles of $\Sigma_{\rm gas}$ are approximately flat up to $R\sim10~{\rm kpc}$, and most of the [O$_{\rm ~III}$] lines are emitted from gas with similar densities.

\begin{figure*}
  \begin{center}
    \begin{tabular}{c}
    
      \begin{minipage}{0.5\hsize}
        \begin{center}
          \includegraphics[bb=0 0 1149 1092, width=\hsize]{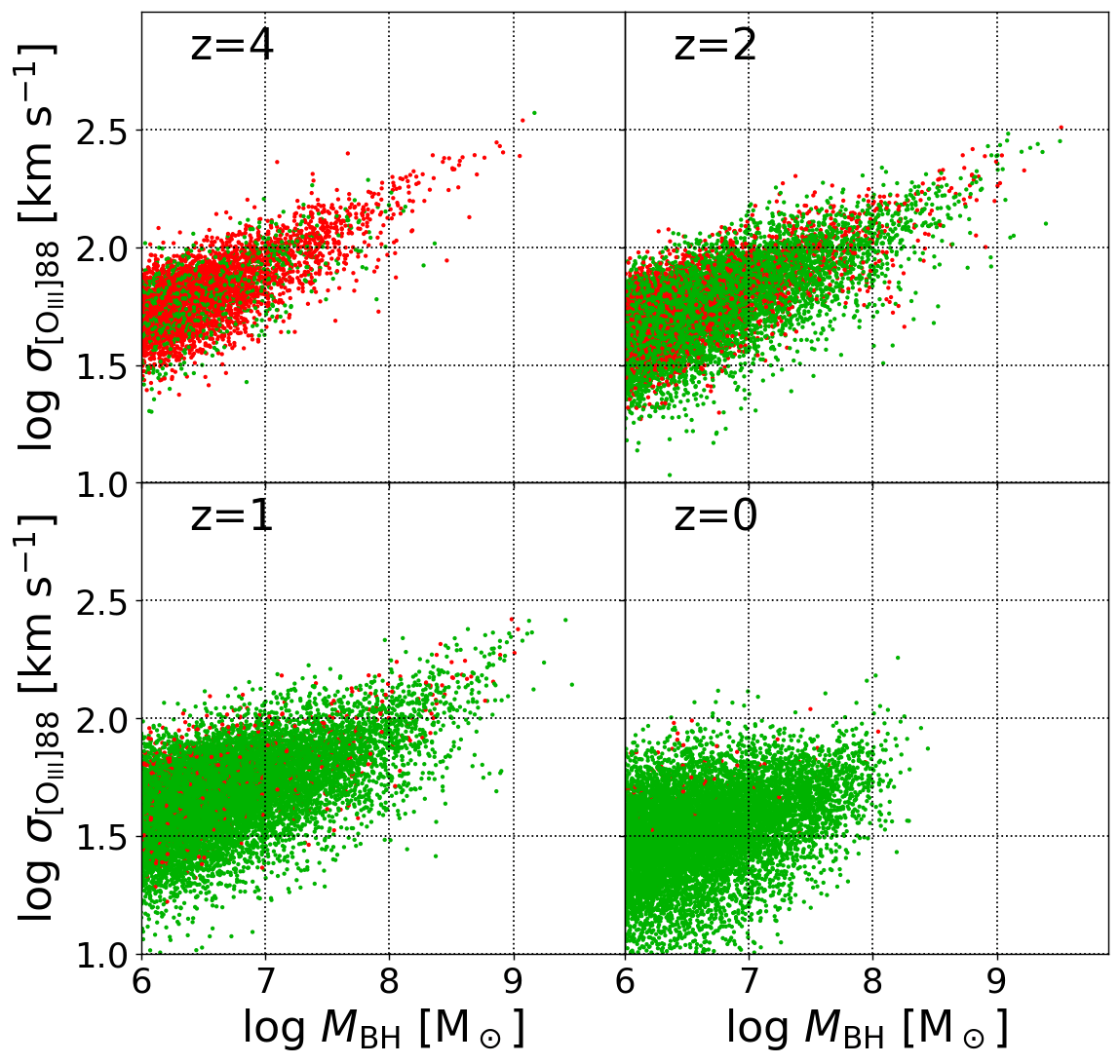}
        \end{center}
      \end{minipage}
      
      \begin{minipage}{0.5\hsize}
        \begin{center}
          \includegraphics[bb=0 0 1149 1092, width=\hsize]{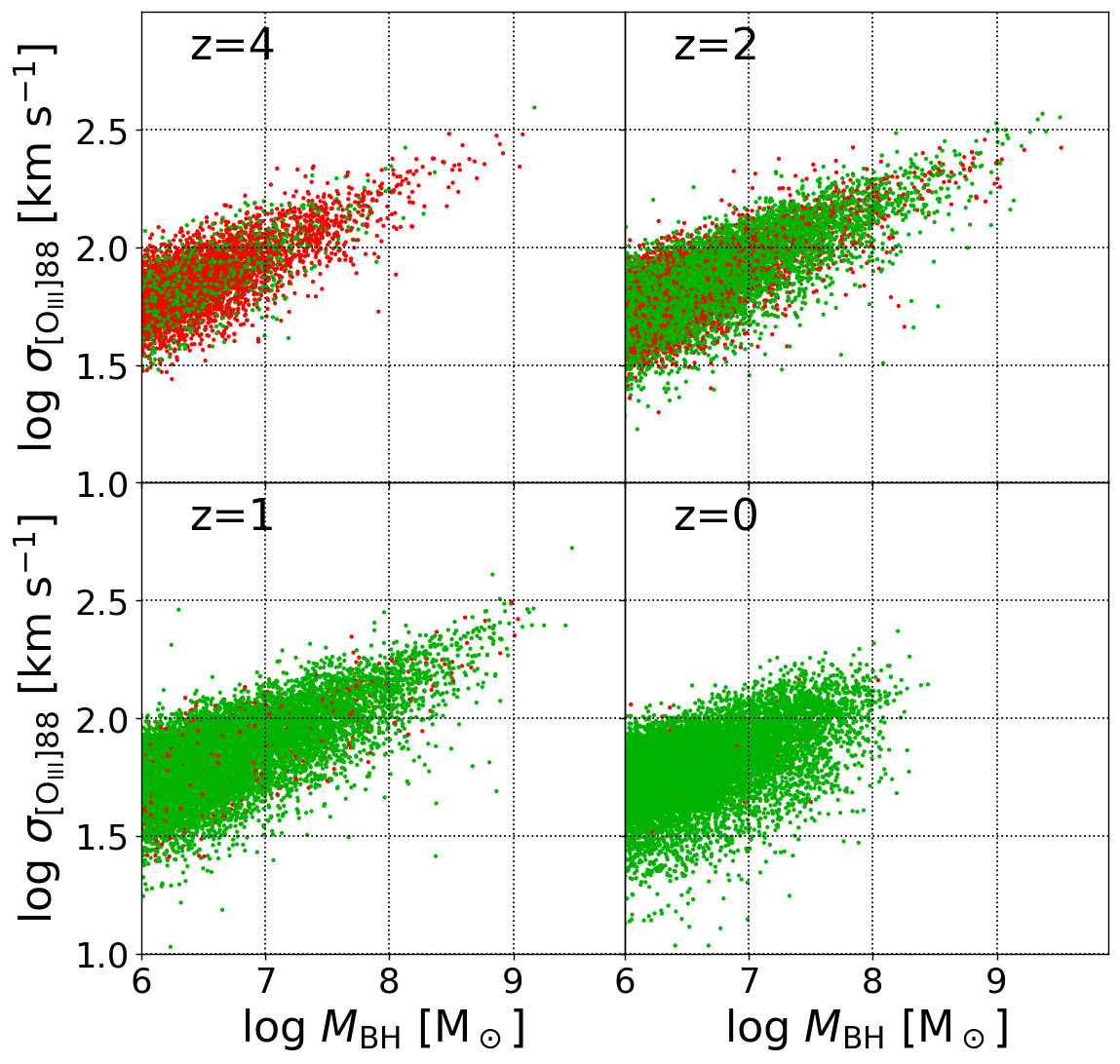}
        \end{center}
      \end{minipage}
    
    \end{tabular}
    \caption{Same as Fig. \ref{BH_VD_TNG} but for the Illustris simulation.}
    \label{VD_MBH_Ill}
  \end{center}
\end{figure*}

Fig. \ref{VD_MBH_Ill} shows the correlations between $M_{\rm BH}$ and $\sigma_{\rm [O_{III}]88}$ for the star-forming galaxies in Illustris. In the distributions, there is no clear segregation between galaxies in the high- and low-accretion modes. In comparing the left and right panels, the values of $\sigma_{\rm [O_{III}]88}$ hardly depend on aperture size although the VDs are slightly higher in the case of the galaxy-integrated apertures at redshift $z=0$. The VDs weighted by [O$_{\rm ~III}]88$ are clearly correlated with $M_{\rm BH}$ although scatters are quite large at $z=0$. It is noteworthy that, in the left set of panels, $\sigma_{\rm [O_{III}]88}$ measured in the 3-kpc apertures also correlate with $M_{\rm BH}$ even in the range of $M_{\rm BH}\gtrsim10^{8.5}~{\rm M_\odot}$ in Illustris. This is clearly different from the result of TNG shown in Fig. \ref{VD_MBH_TNG} where the correlation disappears above the critical mass $M_{\rm BH}\gtrsim10^{8.5}~{\rm M_\odot}$. It implies that measuring $\sigma_{\rm [O_{III}]88}$ with small observational apertures for galaxies hosting massive BHs may be useful to study the impact of AGN on their central gas, and the absence of the correlation at high $M_{\rm BH}$ may be taken as a sign of the inside-out quenching. 

\begin{figure*}
  \begin{center}
    \begin{tabular}{c}
    
      \begin{minipage}{0.5\hsize}
        \begin{center}
          \includegraphics[bb=0 0 1170 1101, width=\hsize]{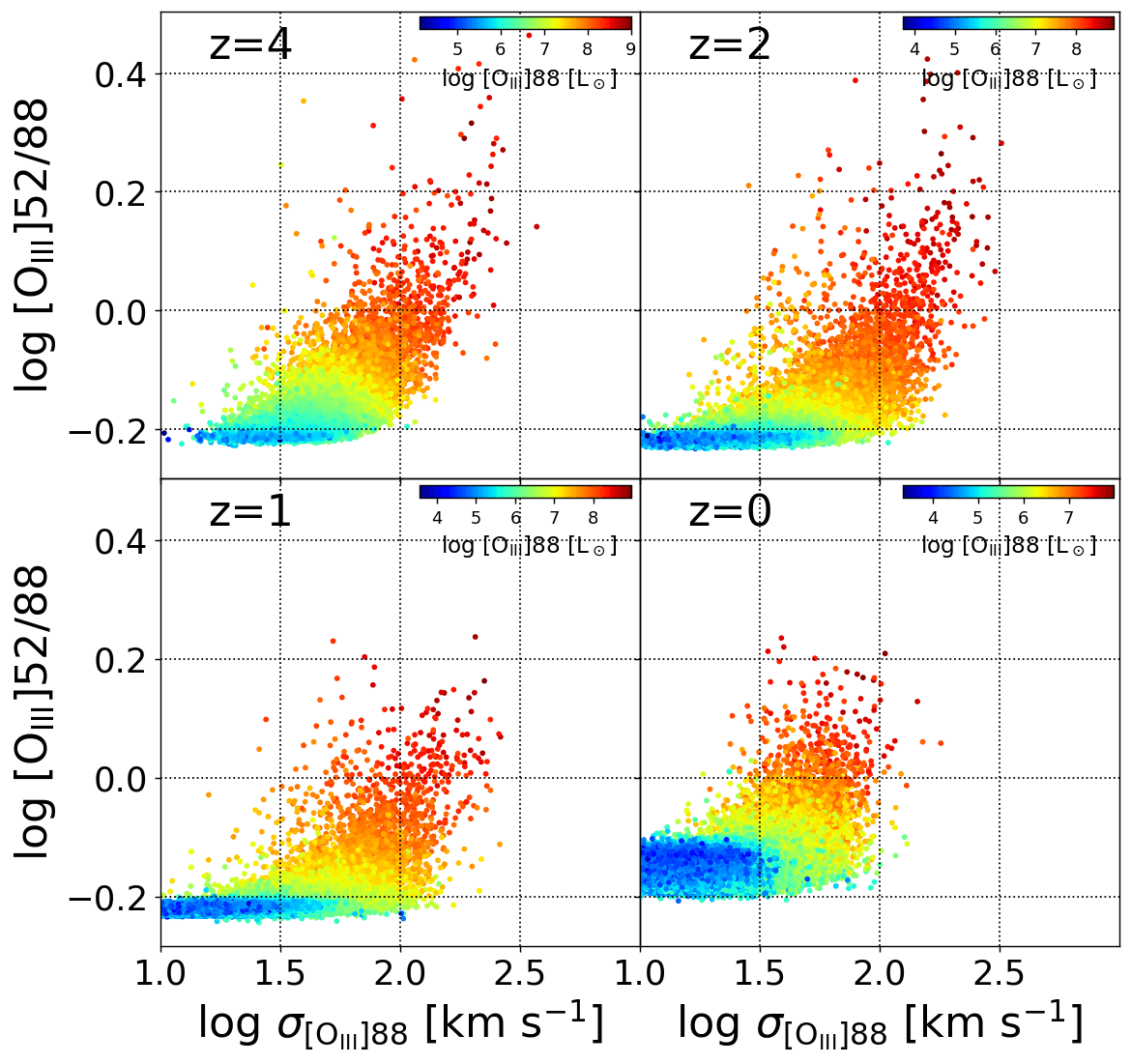}
        \end{center}
      \end{minipage}
      
      \begin{minipage}{0.5\hsize}
        \begin{center}
          \includegraphics[bb=0 0 1170 1101, width=\hsize]{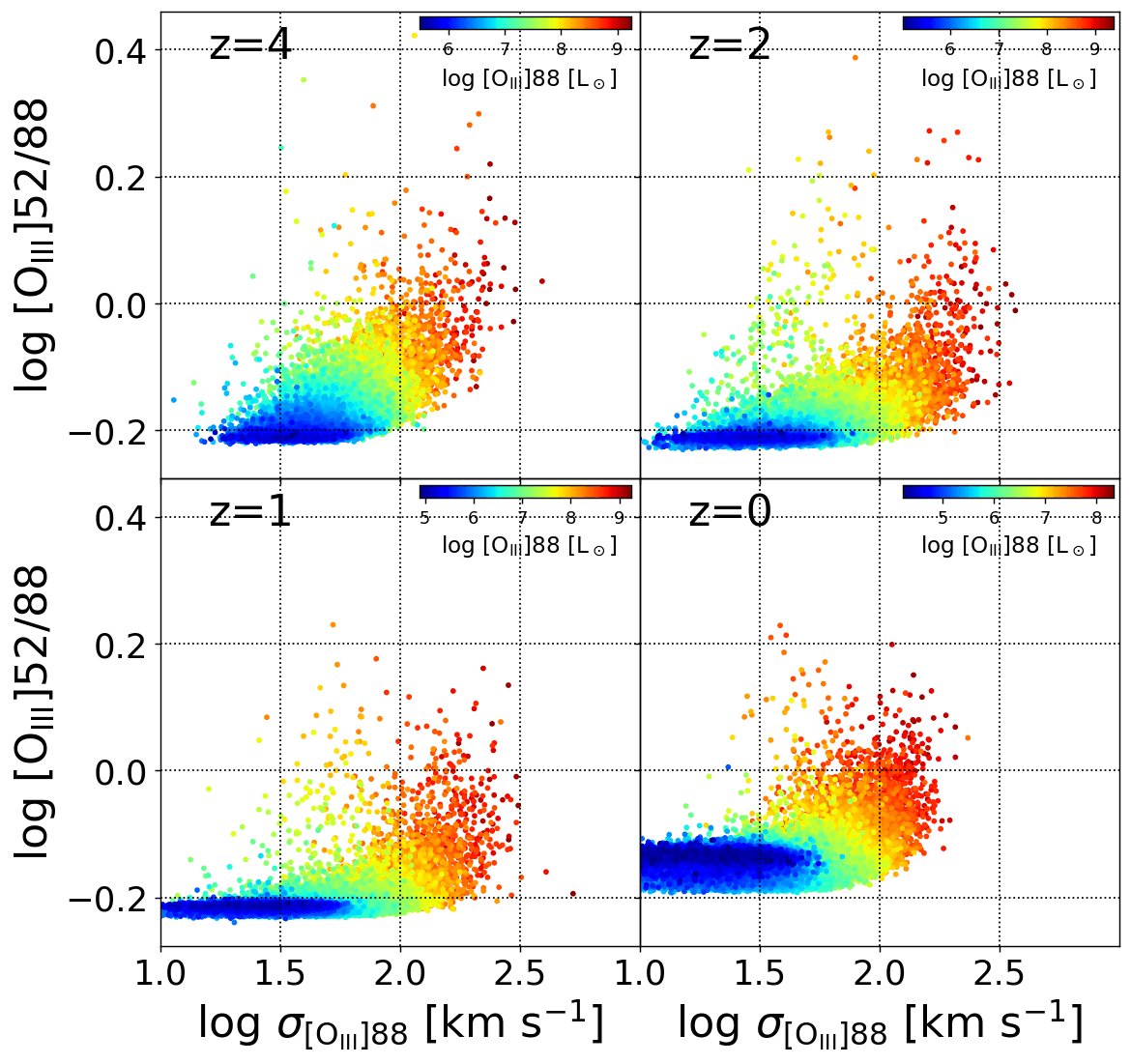}
        \end{center}
      \end{minipage}
    
    \end{tabular}
    \caption{Same as Fig. \ref{VD_MBH_TNG} but for the Illustris simulation.}
    \label{VD_OIII_Ill}
  \end{center}
\end{figure*}
Fig. \ref{VD_OIII_Ill} shows the relationship between $\sigma_{\rm [O_{III}]88}$ and [O$_{\rm ~III}]52/88$. Due to the correlations between $\sigma_{\rm [O_{III}]88}$ and $M_{\rm BH}$, Fig. \ref{VD_OIII_Ill} appears similar to Fig. \ref{OIII_Ill}, where the line ratios increase with the VDs without the drop of [O$_{\rm ~III}]52/88$ at any $\sigma_{\rm [O_{III}]88}$. Thus, if the BH feedback model of Illustris is accurate, measurements of $M_{\rm BH}$ can be replaced with those of $\sigma_{\rm [O_{III}]88}$. In comparing with the result of TNG (Fig. \ref{VD_MBH_TNG}), the presence/absence of the abrupt drop of [O$_{\rm ~III}]52/88$ at $\sigma_{\rm [O_{III}]88}\sim200~{\rm km~s^{-1}}$ is a noticeable difference that could be used to constrain theoretical models of BH feedback. 

\section{Discussion}
\label{dis}
\subsection{Towards observational diagnosis of BH feedback}
\label{dis_obs}
We have shown good correlation of $M_{\rm BH}$ with $\sigma_{\rm [O_{III}]88}$, but $M_{\rm BH}$ appears to be a better quantity to characterise the impact of the BH feedback on [O$_{\rm ~III}]52/88$ in star-forming galaxies. Suppose that we observationally measure [O$_{\rm ~III}]52/88$ of star-forming galaxies with various BH masses, and suppose that the BH masses are known. From our results for TNG discussed in Section \ref{ResTNG}, we can infer whether there is a characteristic value of $M_{\rm BH}$ to quench star formation around the galactic centres. If the BH feedback is suddenly `switched on' to a highly efficient mode at a certain $M_{\rm BH}$, and efficiently expels the gas from the central region in the manner implemented in TNG, [O$_{\rm ~III}]52/88$ is expected to drop abruptly at the BH mass as shown in Fig. \ref{OIII_TNG}. IllustrisTNG predicts the drop to occur at $M_{\rm BH}\sim10^{8.5}~{\rm M_\odot}$. However, if the efficiency of the BH feedback depends on properties other than $M_{\rm BH}$, the variation of [O$_{\rm ~III}]52/88$ would not be clear and the line ratio may change gradually with $M_{\rm BH}$. If the BH feedback is not efficient enough to drastically reduce the central gas density as in Illustris, [O$_{\rm ~III}]52/88$ would not decrease with $M_{\rm BH}$. In fact, as shown in Fig. \ref{OIII_Ill}, Illustris predicts the line ratio to continuously increase with $M_{\rm BH}$ for the star-forming galaxies, although with large scatter. Based on the results, we expect that observations of galaxies at redshifts $z\lesssim3$ may allow us to make direct comparison with the simulations.

If we assume that the BH feedback expels the central gas as reproduced in IllustrisTNG, a low (high) line ratio of [O$_{\rm ~III}]52/88$ implies the presence (absence) of a BH massive enough to trigger quenching the star formation activity in the galaxy. Unfortunately, for this method to work in practice, one needs a prior knowledge on the characteristic $M_{\rm BH}$ for quenching and on weather $M_{\rm BH}$ is a primary quantity 
to cause the inside-out depletion of gas.

In IllustrisTNG, the critical value of $M_{\rm BH}\sim10^{8.5}~{\rm M_\odot}$ varies little with redshift as shown in Figs .\ref{mbh-rho_tng} and \ref{OIII_TNG} \citep[see also][]{wsh:17,tbp:20}. At $z\gtrsim4$, the central BHs are still small in mass, and none of them is in the low-accretion modes in TNG. Massive BHs with $M_{\rm BH}\gtrsim10^{9}~{\rm M_\odot}$ have, however, been actually observed in galaxies at $z\gtrsim6$ \citep[][and references therein]{ms:12}. It would be interesting to observe [O$_{\rm ~III}]52/88$ for such high-redshift galaxies hosting massive BHs. Low line ratios, if measured, are indicative of low gas densities in the galaxies, which in turn
suggest early phases of the inside-out quenching processes driven by BHs. The [O$_{\rm ~III}$] lines emitted at $z\gtrsim6$ are redshifted to submillimetre wavelengths, and ALMA can detect the line emission. The capability of ALMA for detecting the [O$_{\rm ~III}$] lines from such high-redshift galaxies have already been suggested from theoretical models \citep[e.g.][]{ist:14,ayn:20}, and ALMA has indeed observed [O$_{\rm ~III}]88$ of galaxies at $z\gtrsim6$ \citep[e.g.][]{itm:16,hlm:18,nbd:19}. An analytic model of \citet{ylp:21} suggests that the high-redshift galaxy SXDF-NB1006-2 at $z=7.2$ is a promising first target for [O$_{\rm ~III}]52$ measurements. Clearly, using [O$_{\rm ~III}]52/88$ as diagonostics of BH feedback is realistic with ALMA.

The fine structure line diagnostics we propose here is not limited to [O$_{\rm ~III}]52/88$. The line ratio of [N$_{\rm ~II}$] $122$ to $205~{\rm \mu m}$ is also a promising density tracer \citep[e.g.][]{zlx:16,dgi:20}, and the two lines can be used essentially in the same way. ALMA can detect the [N$_{\rm ~II}$] lines from high-redshift galaxies \citep[e.g.][]{nbd:19}. The [O$_{\rm ~III}$] line ratios significantly increase with density $n_{\rm e}$ in the range 10 $\lesssim n_{\rm e} \lesssim 10^4~{\rm cm^{-3}}$, where the logarithmic gradient $[d\ln({\rm [O_{\rm ~III}]52/88})/d\ln n_{\rm e}] >0.1$. The [N$_{\rm ~II}]122/205$ ratio is a sensitive probe at relatively lower densities from $n_{\rm e}\sim1$ to $\sim10^3~{\rm cm^{-3}}$ with $[d\ln({\rm [N_{\rm ~II}]122/205})/d\ln n_{\rm e}] >0.1$ \citep[e.g.][]{dgi:20}.

For a recently discovered quasar at $z=7.54$ with $M_{\rm BH}=8\times10^{8}~{\rm M_\odot}$, \citet{nbd:19} report detection of emission lines of various ions and molecules with ALMA. Although their observations lack [O$_{\rm ~III}]52$, they observe the [N$_{\rm ~II}]$ lines at $205$ and $122~{\rm \mu m}$ with a large aperture diameter of $13~{\rm kpc}$, and place a lower limit of the line ratio to be [N$_{\rm ~II}]122/205>4.4$. The line ratio indicate the local density of the emitting region $n_{\rm e}\gtrsim180~{\rm cm^{-3}}$. It should be noted that the galaxy-integrated line ratio cannot be directly converted to the local density in this case. Using their model for photodissociation regions, they also derive an upper limit of the local density for the quasar to be $n_{\rm H}\lesssim5\times10^4~{\rm cm^{-3}}$. Since there are no such massive BHs at $z\gtrsim6$ in TNG or Illustris simulations, we cannot perform direct comparison. Instead, let us compare the observation with the simulated galaxies at $z\lesssim2$ shown in Fig. \ref{mbh-rho_tng}. Even the lower limit of $n_{\rm H}\sim180~{\rm cm^{-3}}$ appears to be significantly higher than the central densities of galaxies with $M_{\rm BH}\gtrsim10^{8}~{\rm M_\odot}$ in IllustrisTNG. On the assumption that the physical conditions are similar, this may imply that inside-out quenching has not begun in the quasar host galaxy at $z=7.54$. \citet{hit:19} observe two quasars with $M_{\rm BH}\sim10^9~{\rm M_\odot}$ at $z=6.0$ using ALMA and detect bright [O$_{\rm ~III}]88$ emission. They also find physically more extended emission than their dust continua, suggesting large-scale star formation activity. It would be interesting to investigate whether the extended [O$_{\rm ~III}]88$ is relevant to the inside-out quenching in the quasars by observing [O$_{\rm ~III}$] and the [N$_{\rm ~II}$] lines in the future.

For physically extended objects such as high-redshift galaxies, interferometers such as ALMA often take a strategy to enhance the sensitivity at the sacrifice of spatial resolution by shrinking the arrangement of the telescope arrays. Then the central regions of galaxies are not resolved. It is encouraging that the correlations between [O$_{\rm ~III}]52/88$ and $M_{\rm BH}$ are hardly different between the cases with the 3-kpc and galaxy-integrated apertures (Figs. \ref{OIII_TNG} and \ref{OIII_Ill}). The diagnostics we propose here do not require a high spatial resolution if $M_{\rm BH}$ is known. Fortunately, as shown in Figs. \ref{VD_MBH_TNG} and \ref{VD_MBH_Ill}, $M_{\rm BH}$ may be replaced with the second-moments $\sigma_{\rm [O_{III}]88}$ if the aperture size is large enough to cover the entire star-forming regions.

\subsection{Speculation on the BH models from previous observations}
\label{dis_expect}
Some green-valley galaxies have been observed to have lower densities of molecular gas. They also show low SFRs in their central regions, surrounded by outer disc regions with higher densities and SFRs (\citealt{lbp:17,bbm:20}, see also \citealt{elt:20}). \citet{kcs:21} classify local galaxies into various quenching stages according to the morphology of the H$\alpha$ emission map, and find that the `centrally-quiescent' galaxies reside in the green valley. They argue that their sample galaxies that are on the way to `red-sequence' generally begin to cease their star formation from the centres. The observations and the interpretation are consistent with the inside-out quenching process, which is possibly driven by the central BHs. It is also consistent with the cavity and the reverse trend in $\Sigma_{\rm gas}$ shown in Figs. \ref{CObrightest} and \ref{gasdensity} for TNG. It is not clear, however, whether the transition is closely related with the BH mass; it may well be that all galaxies evolve similarly toward  completely quenched states.

\citet{tbp:20} give similar prediction to ours. Using IllustrisTNG, they show that galaxies at $z=0$ are suddenly quenched by the kinematic feedback of BHs at $M_{\rm BH}\sim10^{8.2}~{\rm M_\odot}$, and abrupt transition from star-forming to quiescent galaxies is seen in diagrams of $M_{\rm BH}$, $M_{\rm star}$ and specific SFR. They compare the simulated and observed galaxies at $z=0$ using the diagrams, and find that, for observed galaxies, the transition from star-forming to quiescent states occurs roughly at $M_{\rm BH}\sim10^8~{\rm M_\odot}$. However, the transition is not as sharp as in TNG, and star-forming and quiescent galaxies coexist in a wide range of $M_{\rm BH}\sim10^7$--$10^9~{\rm M_\odot}$. Their result does not point to a clear critical $M_{\rm BH}$ for the transition in the observed galaxies. The quenching process by BH feedback is thought to depend not only on $M_{\rm BH}$ but also on other physical parameters. Accordingly, we speculate that, at least for local galaxies, the line ratios of [O$_{\rm ~III}]52/88$ may not indicate the abrupt decrease at $M_{\rm BH}\sim10^{8.5}~{\rm M_\odot}$. However, we note that our samples taken from the simulations are star-forming galaxies. Our result for TNG shows that the drop of [O$_{\rm ~III}]52/88$ is seen in the star-forming galaxies where their massive BHs have started quenching the galaxies. Because the line ratio traces local gas density rather than the total star formation rate, the diagnostics with the line ratios could more directly deduce in which galaxies the BH quenching operates.

\subsection{[O$_{\rm ~III}$] emission from AGN}
\label{AGNradiation}
Our model considers [O$_{\rm ~III}$] emission from H$_{\rm ~II}$ regions whose energy sources are young star clusters. A luminous AGN can also radiate high-energy photons and thus can be an energy source for the [O$_{\rm ~III}$] emission of gas in the galactic centre. This means that our model can possibly underestimate the absolute luminosities of the [O$_{\rm ~III}$] lines. Since our main conclusions are based on the relative strength of the [O$_{\rm ~III}$] lines, they may not be significantly affected by the influence by an AGN on their absolute luminosities. We note that AGNs can enhance the overall [O$_{\rm ~III}$] emission from inner regions in the galaxy. In this case, the integrated line ratio is more contributed from the emission from the central density, which strongly reflects the efficiency of BH quenching. Although hard photons emitted from an AGN likely increases the ionization parameter $U_{\rm in}$ of the ISM, the line ratio hardly depends on the ionization parameter. In Appendix \ref{app2}, we show that the contribution of AGN radiation to the total [O$_{\rm ~III}$] emission is expected to be insignificant in most of the simulated galaxies. We consider that our results for [O$_{\rm ~III}]52/88$ are not compromised by the uncertainty in the actual $U_{\rm in}$. 

\section{Conclusions and Summary}
\label{con}
The relative strength of a specific pair of line emission such as [O$_{\rm ~III}]52$ and [O$_{\rm ~III}]88$ is used to estimate the local gas density of star-forming regions where the lines are emitted. By postprocessing the outputs of the cosmological simulations of IllustrisTNG and Illustris, we model emission of the [O$_{\rm ~III}$] lines from the simulated galaxies in a large cosmological volume from $z=4$ to $0$.

In galaxies hosting massive BHs, if the BH feedback and star-formation quenching proceeds efficiently in an "inside-out" fashion, the central gas densities are significantly lowered. 
If the BH quenching is inefficient, the gas densities are expected to be higher towards the centres. Such a difference of the central gas densities can be probed by the line ratio of [O$_{\rm ~III}]52/88$ measured in the galaxies. Detecting and identifying both the lines can help extracting the information of the 
physical properties of galactic centres affected by the BH feedback.

For star-forming galaxies in IllustrisTNG, we find that [O$_{\rm ~III}]52/88$ weakly increases with $M_{\rm BH}$ with a large scatter but abruptly decreases at $M_{\rm BH}\sim10^{8.5}~{\rm M_\odot}$. The drop of the line ratios is due to the transition from high- to low-accretion modes of the BH feedback. The line ratios do not significantly depend on aperture size to integrate the [O$_{\rm ~III}$] lines. In addition, we find that the second-moments of [O$_{\rm ~III}]88$ correlate with $M_{\rm BH}$ if the aperture sizes are sufficently large. Hence, in the diagrams of [O$_{\rm ~III}]52/88$ and $\sigma_{\rm [O_{III}]88}$, the abrupt drop of [O$_{\rm ~III}]52/88$ is still seen at a VD of $\sigma_{\rm [O_{III}]88}\sim200~{\rm km~s^{-1}}$. However, if small apertures are applied to measure $\sigma_{\rm [O_{III}]88}$, the correlation between $\sigma_{\rm [O_{III}]88}$ and $M_{\rm BH}$ disappears above the critical mass $M_{\rm BH}\sim10^{8.5}~{\rm M_\odot}$ due to the strong feedback of the massive BHs in TNG.

Models of the BH feedback are, however, largely different among simulations. In Illustris, we find that the line ratios continuously increase with $M_{\rm BH}$, and the drop of [O$_{\rm ~III}]52/88$ does not appear at any $M_{\rm BH}$. This is because the transition from high- to low-accretion modes of the feedback model in Illustris does not significantly affect the central gas densities in star-forming galaxies. The second-moments $\sigma_{\rm [O_{III}]88}$ correlate with $M_{\rm BH}$ regardless of the feedback modes of BHs and the aperture sizes. Thus, the correlations of [O$_{\rm ~III}]52/88$ with $M_{\rm BH}$ and $\sigma_{\rm [O_{III}]88}$ among star-forming galaxies are expected to reflect the efficiency of the BH feedback.

Based on the simulation results, we propose that observing [O$_{\rm ~III}]52/88$ for star-forming galaxies with massive BHs can be used as diagnostics for accuracy of BH models implemented in simulations. If [O$_{\rm ~III}]52/88$ is low in a galaxy with a massive BH, the galaxy may be at an early stage of the inside-out quenching by the BH. We have shown that [O$_{\rm ~III}]52/88$ hardly depends of aperture size, and thus observations do not need to resolve the central regions of galaxies to apply our diagnostics, even if the aperture size is as large as covering the entire galaxies. ALMA is capable of observing [O$_{\rm ~III}]52/88$ and other line ratios such as [N$_{\rm ~II}]122/205$ for high-redshift galaxies at $z\gtrsim6$. Measuring the line ratios for distant galaxies with massive BHs will provide us with important clues to know on how the BH feedback operates in the galaxies.

\section*{Acknowledgements}
We thank Takuya Hashimoto and Ken-ichi Tadaki for their fascinating discussion and helpful suggestion. This study was supported by National Astronomical Observatory of Japan (NAOJ) ALMA Scientific Research Grant Number 2019-11A. HY receives the funding from Grant-in-Aid for Scientific Research (No. 17H04827 and 20H04724) from the Japan Society for the Promotion of Science (JSPS). The numerical computations presented in this paper were carried out on the analysis servers and the general-purpose PC cluster at Center for Computational Astrophysics, NAOJ.

\section*{Data availability}
The data underlying this article will be shared on reasonable request to the corresponding author.



\bibliographystyle{mnras}



%
\appendix

\section{Star-formation main sequence}
\label{App1}
In Section \ref{SFMS}, we use the method of \citet{dpn:19} to define an SFMS and to sample star-forming galaxies in each snapshot. Although \citet{dpn:19} have proposed various methods to determine an SFMS, here we describe the one we use in this study. We assume a power-low relation between $M_{\rm star}$ and SFRs of galaxies,
\begin{equation}
\log\dot{M}_{\rm star} = A\log M_{\rm star}+B,
\label{linearSFMS}
\end{equation}
where $M_{\rm star}$ and $\dot{M}_{\rm star}$ are the total stellar mass and instantaneous SFR of a gravitationally bound structure detected with {\sc SUBFIND}. We divide galaxies into stellar-mass bins with intervals of $0.2~{\rm dex}$ in the range from $M_{\rm star}=10^9$ to $10^{10.2}~{\rm M_\odot}$ and compute the median value $Q_{\rm SFR}$ and standard deviation $\sigma_{\rm SFR}$ of logarithmic SFRs in each bin. We then exclude the galaxies whose SFRs are lower than $Q_{\rm SFR}-2.5\sigma_{\rm SFR}$, re-compute $Q_{\rm SFR}$ and $\sigma_{\rm SFR}$ and iterate this procedure until convergence. Using the converged $Q_{\rm SFR}$ of the bins, we eventually determine the constants $A$ and $B$ in equation (\ref{linearSFMS}) by the least squares method. We set the boundary between star-forming and quenched galaxies to be at $2.5\overline{\sigma}_{\rm SFR}$ below equation (\ref{linearSFMS}) for all stellar masses, where $\overline{\sigma}_{\rm SFR}$ is the median of $\sigma_{\rm SFR}$. 

\begin{figure}
  \includegraphics[bb=0 0 1218 1106, width=\hsize]{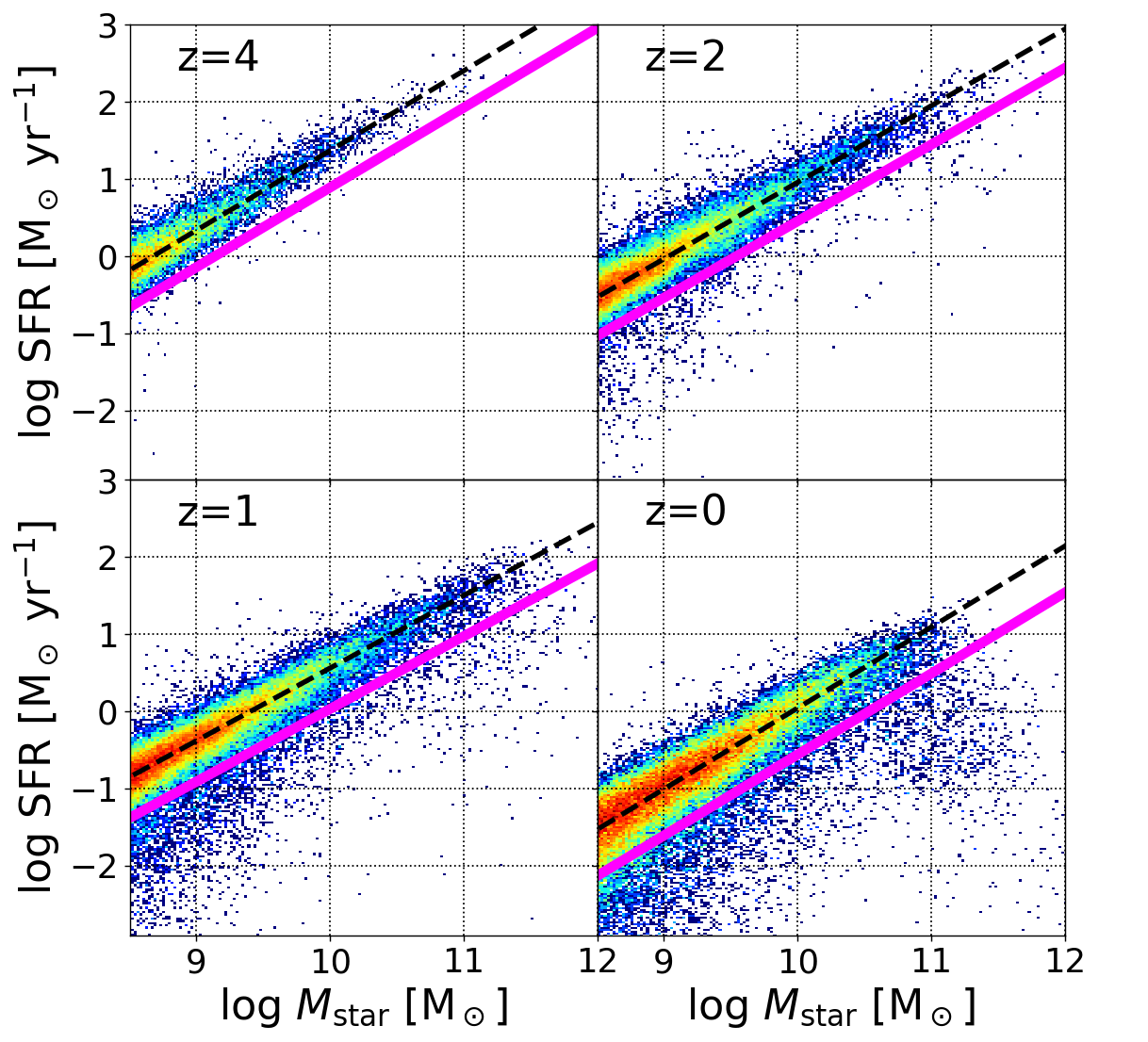}
  \caption{Same as Fig. \ref{SFMSTNG} but for the Illustris simulation.}
  \label{SFMSILL}
\end{figure}
Fig. \ref{SFMSILL} shows the same result as Fig. \ref{SFMSTNG} but for Illustris. Generally, the offset of the boundary from the SFMS, $2.5\overline{\sigma}_{\rm SFR}$, decreases with redshift and is smaller than $1~{\rm dex}$ in all snapshots in both simulations. \citet{dpn:19,dpn:20} have argued the dependence of estimated fractions of quenched galaxies and relationship with the colour bimodality on their methods to determine SFMSs in IllustrisTNG. 

\section{Influence by radiation from black holes}
\label{app2}
We estimate the contribution from the central AGN as a power source to the galaxy-integrated [O$_{\rm ~III}$] emission.
It is difficult to accurately compute the line emission since we need to know the emergent SED of an AGN which generally has unresolved small-scale structures such as torus. We thus resort to estimating a bolometric luminosity of the BH and compare it with the total bolometric luminosity of stars that form H$_{\rm ~II}$ regions in a galaxy.

We follow a popular model of AGN that assumes the bolometric luminosity is given by
\begin{align}
L_{\rm BH} = 
\begin{cases} 
\epsilon_r\dot{M}_{\rm BH} c^2 \quad &\mbox{for} \quad \dot{M}_{\rm BH} \ge 0.1\dot{M}_{\rm Edd} , \\
 10\left(\frac{\dot{M}_{\rm BH}}{ \dot{M}_{\rm Edd}}\right)^2\epsilon_r\dot{M}_{\rm Edd}c^2 \quad &\mbox{for} \quad \dot{M}_{\rm BH} < 0.1\dot{M}_{\rm Edd},
 \end{cases}
\label{LbolBH}
\end{align}
where $\epsilon_r=0.2$ which is consistent with the value used in the simulations \citep{css:05,hds:14,wsp:18}. The bolometric luminosities of star clusters are computed with {\sc P\'{E}GASE.2} in the same way as in our model of the [O$_{\rm ~III}$] lines (see Section \ref{method}).

\begin{figure}
    \includegraphics[bb=0 0 1106 1106, width=\hsize]{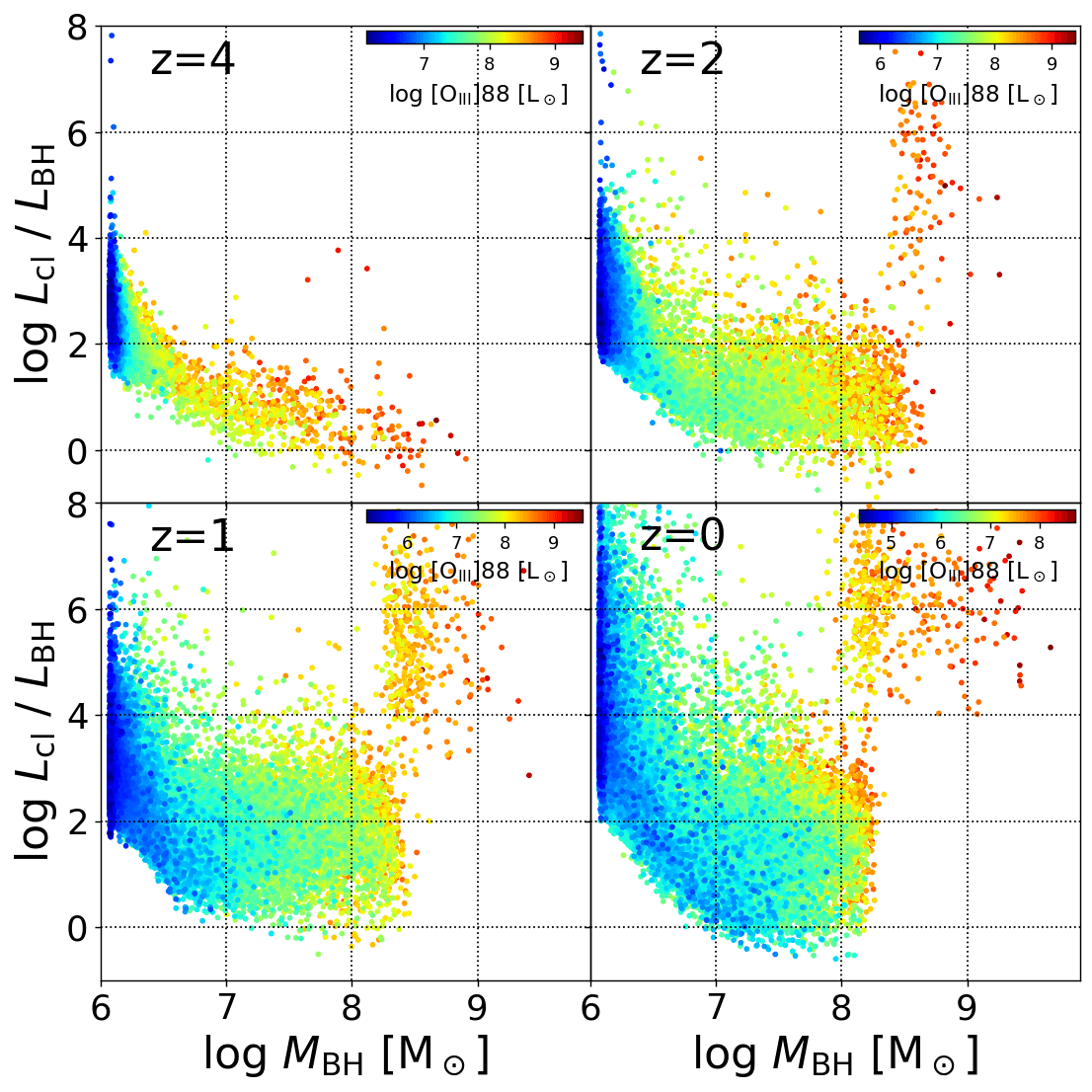}
    \caption{The total bolometric luminosity of star clusters, $L_{\rm cl}$, normalized by that of the most massive BH, $L_{\rm BH}$, as a function of $M_{\rm BH}$. We use the galaxy sample in IllustrisTNG. The colours of the plotted dots indicate the galaxy-integrated [O$_{\rm ~III}]88$ luminosities.}
    \label{Lbol_TNG}
\end{figure}
\begin{figure}
    \includegraphics[bb=0 0 1106 1106, width=\hsize]{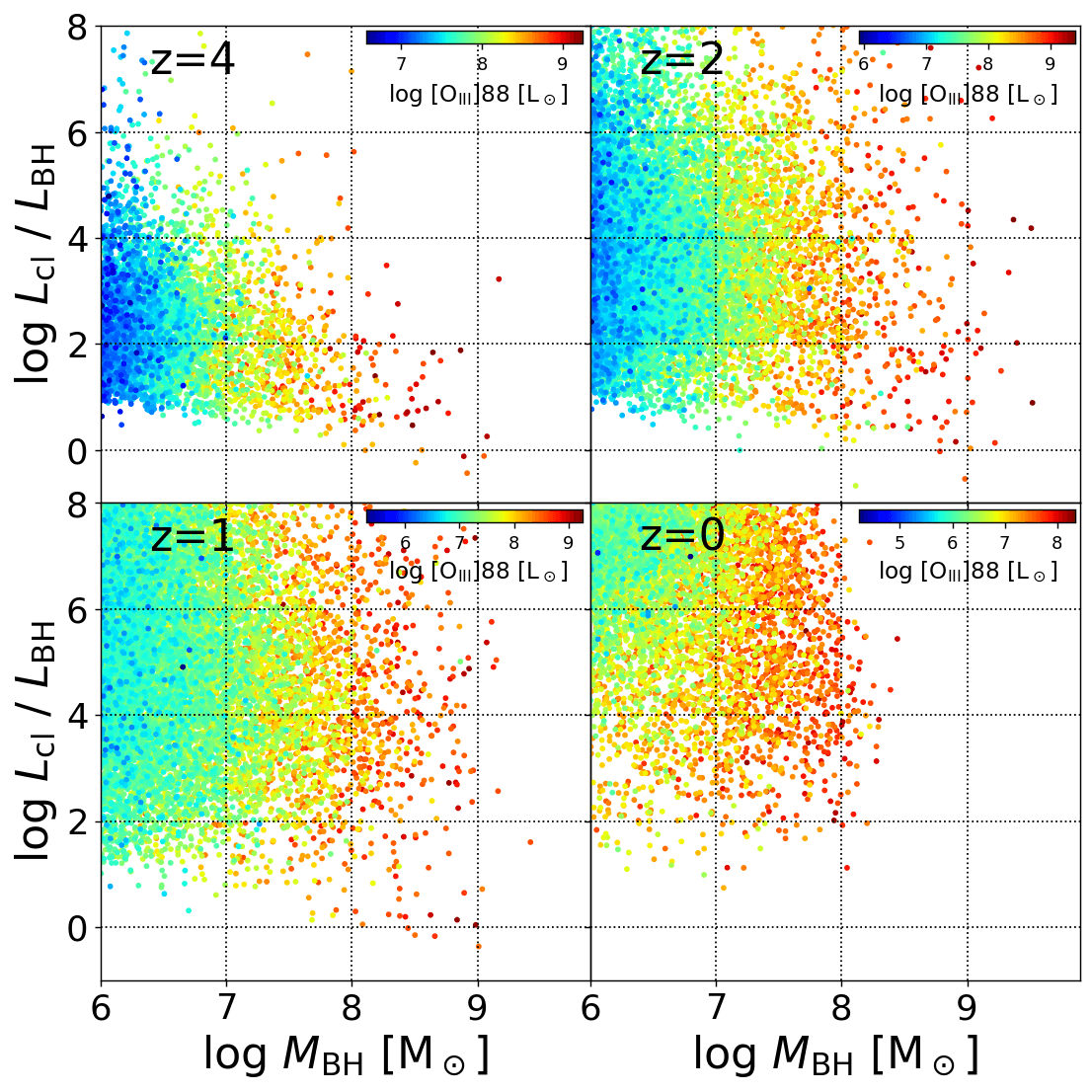}
    \caption{Same as Fig. \ref{Lbol_TNG} but for Illustris.}
    \label{Lbol_Ill}
\end{figure}
For the star-forming galaxies in IllustrisTNG and Illustris, Figs \ref{Lbol_TNG} and \ref{Lbol_Ill} show the ratios of the galaxy-integrated bolometric luminosities of star clusters with respect to $L_{\rm BH}$ of the most massive BHs in the galaxies. In both the simulations, most of the galaxies have ratios higher than unity, suggesting that the net radiation energy released from the star clusters is generally larger than that from BHs in the galaxies. We thus consider that BHs in the simulations are less significant sources for the [O$_{\rm ~III}$] lines.

\begin{figure}
    \includegraphics[bb=0 0 766 505, width=\hsize]{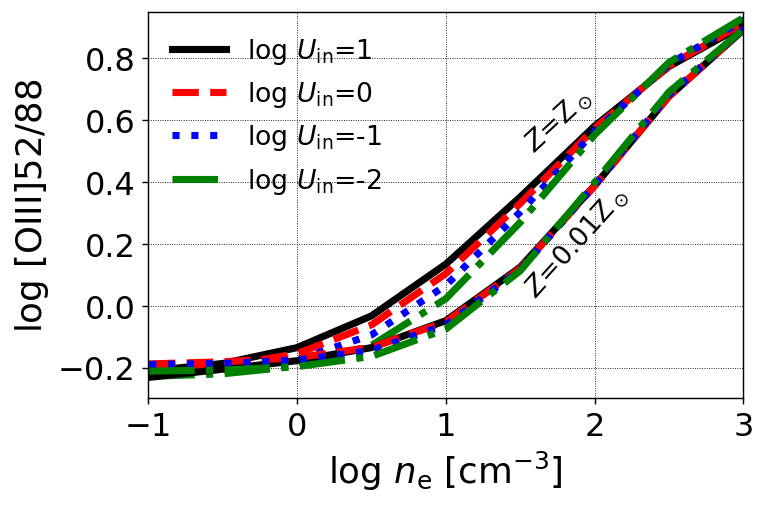}
    \caption{Same as Fig. \ref{LineRatio} but changing $U_{\rm in}$ independently from $n_{\rm e}$. The upper and lower lines delineate the results with $Z={\rm Z_\odot}$ and $Z=0.01{\rm Z_\odot}$.}
    \label{Udep}
\end{figure}

Even if the bolometric luminosity of an AGN is insignificant, it can emit hard photons and may increase the ionization parameters of star-forming regions. To evaluate the influence on [O$_{\rm ~III}]52/88$, we compute the line ratios with our \textsc{Cloudy} model described in Section \ref{method} by manually changing the parameter $U_{\rm in}$. Fig. \ref{Udep} shows the dependence of the computed [O$_{\rm ~III}]52/88$ on $U_{\rm in}$ in the cases with $Z={\rm Z_\odot}$ and $Z=0.01{\rm Z_\odot}$. The line intensity ratio does not strongly depend on $U_{\rm in}$, and therefore our conclusions in the main text based on [O$_{\rm ~III}]52/88$ are not significantly affected by considering the AGN radiation contribution.

\bsp	
\label{lastpage}
\end{document}